\documentclass[11pt]{article}

\usepackage{amsmath,mathtools,amssymb,color,epsfig,cite}
\usepackage{amsfonts}
\usepackage{graphicx}
\usepackage{set	space}	
\usepackage{mathrsfs}
\usepackage{tikz}
\usetikzlibrary{calc,decorations.markings}
\usetikzlibrary{arrows,decorations.pathmorphing}
\usetikzlibrary{positioning}
\usepackage{makecell}
\usepackage{multirow}
\usepackage{enumitem}
\usepackage{float}
\usepackage{braket}

\usepackage{hyperref} 
\hypersetup{
bookmarksnumbered=true,
    colorlinks=true, 
    linkcolor=black, 
    urlcolor=black, 
    citecolor = black,
    linktoc=all 
}

 \tikzset{
 particle/.style={thick,draw=black, postaction={decorate},
     decoration={markings,mark=at position .75 with {\arrow[line width=1pt]{>}}  }},
 aparticle/.style={thick,draw=black, postaction={decorate},
     decoration={markings,mark=at position .5 with  {\arrow[line width=1pt]{<}}  }},
 gluon/.style={decorate, draw=black,
     decoration={coil,aspect=0}}
  }

\makeatletter
\newsavebox{\@brx}
\newcommand{\llangle}[1][]{\savebox{\@brx}{\(\m@th{#1\langle}\)}
  \mathopen{\copy\@brx\kern-0.5\wd\@brx\usebox{\@brx}}}
\newcommand{\rrangle}[1][]{\savebox{\@brx}{\(\m@th{#1\rangle}\)}
  \mathclose{\copy\@brx\kern-0.5\wd\@brx\usebox{\@brx}}}
\makeatother

\usepackage[left=2.5cm,right=2.5cm,top=2.5cm,bottom=2.5cm]{geometry}

\usepackage{amsfonts}

\newcommand{\hoch}[1]{$ ^{#1}$}

\makeatletter
\@addtoreset{equation}{section}
\makeatother

\newcommand{\be}{\begin{equation}}
\newcommand{\ee}{\end{equation}}
\newcommand{\bea}{\setlength\arraycolsep{2pt} \begin{eqnarray}}
\newcommand{\eea}{\end{eqnarray}}
\newcommand{\nn}{\nonumber}

\newcommand{\dbar}{{d\mkern-7mu\mathchar'26\mkern-2mu}}

\newcommand{\intl }{\int\limits  }

\newcommand{\ddd}[1]{\! d#1\ }

\def\biblio{\bibliographystyle{utphys}\bibliography{LambdaPhi4}}

\def\0{{\sst{(0)}}}
\def\1{{\sst{(1)}}}
\def\2{{\sst{(2)}}}
\def\3{{\sst{(3)}}}
\def\4{{\sst{(4)}}}
\def\5{{\sst{(5)}}}
\def\6{{\sst{(6)}}}
\def\7{{\sst{(7)}}}
\def\8{{\sst{(8)}}}
\def\sst#1{{\scriptscriptstyle #1}}

\def\E{{\mathbb E}}

\def\E{\scriptscriptstyle{E}}

\def\0{\scriptscriptstyle{0}}
\def\1{\scriptscriptstyle{1}}
\def\p1{\scriptscriptstyle{(1)}}
\def\2{\scriptscriptstyle{2}}
\def\pp2{\scriptscriptstyle{(2)}}
\def\3{\scriptscriptstyle{3}}
\def\ppp3{\scriptscriptstyle{(3)}}
\def\-{\scriptscriptstyle{-}}
\def\+{\scriptscriptstyle{+}}

\begin{document}
\def\biblio{}

\vspace*{15pt}
\begin{center}
{\Large {\bf ANEC on stress-tensor states \\ in perturbative $\lambda \, \phi^4$ theory} }

\vspace{25pt}
{\bf Teresa Bautista\hoch{1} and Lorenzo Casarin\hoch{2,3}}

\vspace{20pt}

{\hoch{1}\emph{Department of Mathematics, King’s College London,\\The Strand, London WC2R 2LS, UK}\\[1em]
\hoch{2}\emph{Institut f\"ur Theoretische Physik\\Leibniz Universit\"at Hannover
\\
Appelstra\ss{}e 2, 30167 Hannover, Germany}
\\[1em]
\hoch{3}\emph{Max-Planck-Institut f\"ur Gravitationsphysik (Albert Einstein Institut)\\Am M\"uhlenberg  1, D-14476 Potsdam, Germany}
}

 \vspace{25pt}
October 2022

\vspace{30pt}

\underline{Abstract}

\end{center}

\noindent  We evaluate the Average Null Energy Condition (ANEC) on momentum eigenstates generated by the stress tensor  in  perturbative $\lambda \, \phi^4$ and general spacetime dimension. We first compute the norm of the stress-tensor state at second order in $\lambda$; as a by-product of the derivation we obtain the full expression for the stress tensor 2-point function at this order. We then compute the ANEC expectation value to first order in $\lambda$, which  also  depends on the  coupling of the  stress-tensor improvement term $\xi$. 
We study the bounds on these couplings  that follow from the ANEC and unitarity at first order in perturbation theory. 
These bounds are stronger than   unitarity in some regions of coupling space.

 \noindent

\thispagestyle{empty}

\vfill
\hoch{1}teresa.bautista@kcl.ac.uk,\qquad  \hoch{2}lorenzo.casarin@\{itp.uni-hannover.de, aei.mpg.de\}

\pagebreak
\tableofcontents
\vspace{10mm}
\rule{400pt}{0.03pt}
\section{Introduction}

The Average Null Energy Condition (ANEC) states that the integral of the null energy over a complete null worldline is non-negative,
\begin{equation}\label{ANEC-definition} 
\intl_{-\infty}^{+\infty } \ddd{z^-}   \,T_{--}(z) \geq 0\, .
\end{equation}
This is a quantum statement, it  holds true at operatorial level. It is satisfied in free theory \cite{Klinkhammer:1991ki},  it has been shown to hold for interacting unitary QFTs with a nontrivial UV fixed point using field-theoretic methods \cite{Hartman:2016lgu}, and more generally for any unitary QFT using entropy arguments \cite{Faulkner:2016mzt}.

The ANEC \eqref{ANEC-definition} is an inherently  Lorentzian concept. In fact, the central ingredient in the proof of \cite{Hartman:2016lgu}  is causality, which more in general  is   crucial in the analytic conformal bootstrap programme, recently reviewed in \cite{Hartman:2022zik,Bissi:2022mrs}.   
  In this sense, studying the implications of the ANEC lies within the broad program of determining the consequences of causality and unitarity for QFTs.

The ANEC was shown to encode important information about conformal field theories with the derivation of the `conformal collider bounds' \cite{Hofman:2016awc}, which are bounds on conformal anomalies. To derive these,  the ANEC operator is placed at null infinity, and its expectation values are taken on a state \(\ket{\psi}\) which generates some energy excitation,
\begin{equation}\label{efl}
\braket{  E }     =
\frac{1}{\braket{\psi|\psi}}  \lim_{z^+\to \infty} \Big(\frac {z^+}2 \Big)^{d-2} 
\bra{\psi} \intl_{-\infty}^{+\infty }
\ddd{z^-}   \,T_{--}(z) \ket{\psi}
\geq 0 \,.
\end{equation}
This then has the interpretation of the energy flux measured per unit angle in the transverse directions at null infinity, which owing to \eqref{ANEC-definition} has to be non-negative. 

The expectation value  \eqref{efl} is computed from   3-point correlators involving the stress tensor, and their positivity translates into bounds on the quantities which such correlators depend on.
In a CFT, for a  momentum eigenstate generated by the stress tensor itself,
the expectation value in $d=4$ spacetime dimensions depends on the conformal anomalies \(\mathrm a\) and \( \mathrm c\), and the ANEC translates into a lower and an upper bound on their ratio  \(\mathrm a/\mathrm c\). The bounds thus obtained also happen to be optimal, given that the ANEC operator commutes with the momentum operator at null infinity.

The ANEC has also been used to place bounds on conformal dimensions of operators \cite{Cordova:2017dhq,Manenti:2019kbl}, in some cases stronger than the unitary bounds.   
Furthermore,  it has been shown \cite{Hartman:2016lgu,Kravchuk:2018htv} that the ANEC is   the first of a whole family of positivity conditions, which similarly follow from causality and unitarity. These take the form of the positivity of light-ray operators \cite{Kravchuk:2018htv,Kologlu:2019mfz,Caron-Huot:2022eqs}, non-local operators labeled by a continuous spin $J$, for which the $J=2$ operator is precisely the ANEC operator. Their positivity therefore generalises the ANEC to continuous spin.

Given how useful the ANEC has proven to be in the context of CFTs, it is natural to explore its implications for generic QFTs. Since it follows from unitarity and causality, it is tempting to think that the ANEC could encode interesting constraints on RG flows and be related to monotonicity theorems. However, the lack of conformal symmetry  makes it much more difficult to make general statements on the  correlators. It is therefore useful to start by studying a particular example. 

In this paper, we continue the programme initiated in \cite{Bautista:2020bjy} by studying the implications of the ANEC in the particular example of $\lambda \phi^4$ in perturbation theory. This is an interacting theory with a trivial fixed point in $d=4$ dimensions and a   Wilson-Fisher fixed point in $d = 4 - 2 \epsilon$, and it is simple enough to allow one to explicitly compute the expectation value of the ANEC operator at low perturbative orders.  Concretely, here we consider a state generated by the stress-tensor, thereby  following the construction of \cite{Hofman:2016awc} and deriving nontrivial constraints for the parameters of the theory.  For practical purposes we focus on the case  $2<d\leq4$, although some of the results have   a more general range of validity.

The constraints that we obtain from the ANEC depend on the  spacetime dimension, the coupling $\lambda$, the improvement-term coupling  $\xi$, and the energy of the state;  they are trivially satisfied in the free case.  The constraints are similar to the unitarity constraint that follows from demanding positivity of the norm of the state. 
Setting the renormalization scale equal to the energy of the state we obtain bounds for the couplings at such energy.
 The ANEC turns out to be in most cases more stringent than unitarity. The evaluation of the norm of the state and of the ANEC correlator is solid; however, the analysis of the bounds is more speculative given that they follow from the edge of validity of perturbation theory, and  require higher-order corrections to be confirmed.

The outline of the paper is as follows. In section \ref{sec-norm} we compute the norm of the state generated by the stress tensor up to order $\mathcal O(\lambda^2)$, which follows from the Wightman 2-point stress-tensor correlator. In section \ref{sec-numerator} we compute the ANEC expectation value on the same state up to order $\mathcal O(\lambda)$, by first computing the Wightman 3-point correlator of the stress tensor and then turning it into an expectation value of the ANEC operator at null infinity. Finally in section \ref{sec-bounds} we present and discuss the resulting constraints, together with the unitarity constraint following from positivity of the norm of the state.  Several of the intermediate expressions for the correlators are listed in the appendices, including the expression for the full stress-tensor 2-point function to $\mathcal O(\lambda^2)$  and details of the derivation of the Wightman function in momentum space from the Euclidean correlators. In appendix \ref{app-mass}, we compare with the case of the free massive scalar.

\section{Norm of the state}\label{sec-norm}

 Our starting point is the Euclidean action in \(d=4-2\varepsilon\) dimensions
\begin{equation}\label{aaa}
S_\mathrm{\E} = \int \ddd{^dx_\mathrm{\E}} \Big[\frac12 (\partial\phi)^2 + \frac{1}{4!} \lambda \phi^4\Big] \,,
\end{equation}
where   subscripts \(\mathrm E\) (\(\mathrm L\)) indicates Euclidean (Lorentzian)  signature. In the following we will drop them whenever  it is clear from the context.

The Euclidean stress-energy tensor derived from \eqref{aaa}
reads
\begin{equation}\label{EMtensor}
T_{\mu\nu}=
\partial_\mu\phi\,\partial_\nu\phi -\frac{1}{2} \,(\partial\phi)^2\,\delta_{\mu\nu}  -\xi \left(\partial_\mu\partial_\nu -\delta_{\mu\nu}\,\partial^2\right) \phi^2
 -\frac{\lambda}{4!}\,\phi^4\,\delta_{\mu\nu}\,.
\end{equation}
It includes the improvement term with a  real parameter \(\xi\). As we shall confirm with our calculation, its addition  is necessary to construct a renormalizable  energy-momentum tensor   at the quantum level \cite{Freedman:1974ze}. Tracelessness of the stress tensor (when \(\lambda =0\) or \(d =4\)) is achieved when \(\xi = \xi_d := \frac{d-2}{4(d-1)} \).

We want to evaluate the norm of the stress-tensor momentum eigenstate
 \begin{equation}\label{aac-bis}
\ket{\varepsilon\cdot T } = \varepsilon^{\mu\nu}\int \ddd{^dx} e^{- i\,  q \,x^{\0}} T_{\mu\nu}(x) \ket{0} 
\,,
\qquad q>0
\end{equation}
on which we will  evaluate the ANEC operator.
This state has vanishing spatial momentum $p^i=\vec{0}$  and energy \( p^{\0}=q >0\). We introduced also a complex symmetric polarization tensor \(\varepsilon\). Conservation of the stress tensor allows us to consider purely space-like polarization,  \(\varepsilon^{{\0}\mu}=0\). We write the norm as
\begin{equation}\label{N1}
\begin{aligned}
N&= \braket{ \varepsilon\cdot T |     \varepsilon\cdot T  }\\
&= {\varepsilon^*}^{\mu\nu}\varepsilon^{\alpha\beta} \int \ddd{^dx} e^{iqx^{\0}} \braket{T_{\mu\nu}(x)T_{\alpha\beta}(0)}  
=   {\varepsilon^*}^{\mu\nu}\varepsilon^{\alpha\beta}  \langle 
T_{\mu\nu}(q,\vec 0)\,T_{\alpha\beta}(-q,\vec 0) 
\rangle\,,
\end{aligned}
\end{equation}
where the correlator involved is the  Wightman 2-point function.
In the second step we have dropped a factor of the spacetime volume, since it cancels with an analogous contribution from the ANEC 3-point function in the expression for the normalized energy flux \eqref{efl}.

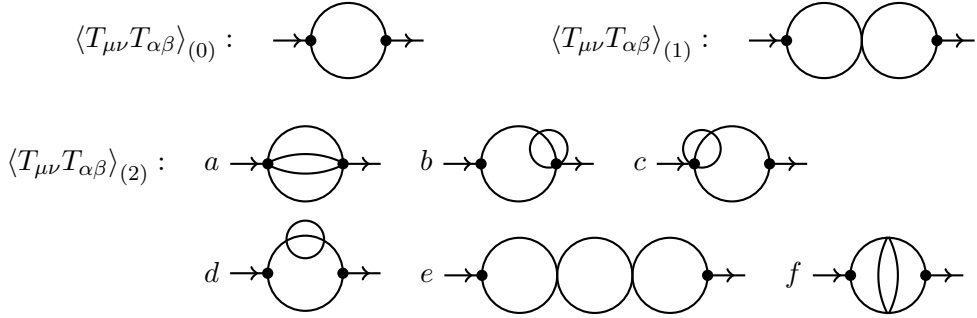
\begin{figure}
\centering
             \( \braket{T_{\mu\nu}  T_{\alpha\beta} }_{ (0)} :\quad \)
 \begin{tikzpicture}[baseline={([yshift=-.5ex]current bounding box.center)},thick]
\draw[particle] (-1,0)--(-0.5,0);
\draw[aparticle] (1,0)--(0.5,0);
\draw (0,0) circle (0.5); 
\node at (-0.5,0)[circle,fill,inner sep = 1.5pt] {};
\node at (0.5,0)[circle,fill,inner sep = 1.5pt] {};
\end{tikzpicture} 
 \(\qquad\qquad \braket{T_{\mu\nu}  T_{\alpha\beta} }_{ (1)} :\quad \)
\begin{tikzpicture}[baseline={([yshift=-.5ex]current bounding box.center)},thick]
\draw[particle] (-1,0)--(-0.5,0);
\draw[aparticle] (2,0)--(1.5,0);
\draw (0,0) circle (0.5); 
\draw (1,0) circle (0.5); 
\node at (-0.5,0)[circle,fill,inner sep = 1.5pt] {};
\node at (1.5,0)[circle,fill,inner sep = 1.5pt] {};
\end{tikzpicture} 
\\[1.5em]
             \( \braket{T_{\mu\nu}  T_{\alpha\beta} }_{ (2)} :\quad \)
             \(a\)
\begin{tikzpicture}[baseline={([yshift=-.5ex]current bounding box.center)},thick]
\draw[particle] (-1,0)--(-0.5,0);
\draw[aparticle] (1,0)--(0.5,0);
\draw (0,0) circle (0.5); 
\node at (-0.5,0)[circle,fill,inner sep = 1.5pt] {};
\node at (0.5,0)[circle,fill,inner sep = 1.5pt] {};
\draw [domain=65:90+25] plot ({1.13*cos(\x)}, {1.13*sin(\x)-1});
\draw [domain=65:90+25] plot ({1.13*cos(\x)}, {-1.13*sin(\x)+1});
\end{tikzpicture} 
             \(\quad b\)
\begin{tikzpicture}[baseline={([yshift=-.5ex]current bounding box.center)},thick]
\draw[particle] (-1,0)--(-0.5,0);
\draw[aparticle] (1,0)--(0.5,0);
\draw (0,0) circle (0.5); 
\node at (-0.5,0)[circle,fill,inner sep = 1.5pt] {};
\node at (0.5,0)[circle,fill,inner sep = 1.5pt] {};
\draw (0.4,0.2) circle (0.25); 
\end{tikzpicture} 
 \(\quad c\)
\begin{tikzpicture}[baseline={([yshift=-.5ex]current bounding box.center)},thick]
\draw[particle] (-1,0)--(-0.5,0);
\draw[aparticle] (1,0)--(0.5,0);
\draw (0,0) circle (0.5); 
\node at (-0.5,0)[circle,fill,inner sep = 1.5pt] {};
\node at (0.5,0)[circle,fill,inner sep = 1.5pt] {};
\draw (-0.4,0.2) circle (0.25); 
\end{tikzpicture}
\hspace{3cm}

\vspace*{0.5em}
\hspace{1.45cm}
 \(  d\)
\begin{tikzpicture}[baseline={([yshift=-6pt]current bounding box.center)},thick]
\draw[particle] (-1,0)--(-0.5,0);
\draw[aparticle] (1,0)--(0.5,0);
\draw (0,0) circle (0.5); 
\node at (-0.5,0)[circle,fill,inner sep = 1.5pt] {};
\node at (0.5,0)[circle,fill,inner sep = 1.5pt] {};
\draw (0,0.45) circle (0.25); 
\end{tikzpicture} 
 \(\quad e\)
\begin{tikzpicture}[baseline={([yshift=-.5ex]current bounding box.center)},thick]
\draw[particle] (-1,0)--(-0.5,0);
\draw[aparticle] (3,0)--(2.5,0);
\draw (0,0) circle (0.5); 
\draw (1,0) circle (0.5); 
\draw (2,0) circle (0.5); 
\node at (-0.5,0)[circle,fill,inner sep = 1.5pt] {};
\node at (2.5,0)[circle,fill,inner sep = 1.5pt] {};
\end{tikzpicture} 
 \(\quad f\)
\begin{tikzpicture}[baseline={([yshift=-.5ex]current bounding box.center)},thick]
\draw[particle] (-1,0)--(-0.5,0);
\draw[aparticle] (1,0)--(0.5,0);
\draw (0,0) circle (0.5); 
\node at (-0.5,0)[circle,fill,inner sep = 1.5pt] {};
\node at (0.5,0)[circle,fill,inner sep = 1.5pt] {};
\draw [domain=63-90:27] plot ({1.13*cos(\x)-1}, {1.13*sin(\x)});
\draw [domain=63+90:180+27] plot ({1.13*cos(\x)+1}, {-1.13*sin(\x)});
\end{tikzpicture} 

\caption{Diagrams for the 2-point function. The thick dot represents a  stress tensor insertion. The left one has indices \(\mu\nu\); the  right one  \(\alpha\beta\). The arrow represents the flow of the external momentum \(p\).}
\label{fig:TTdiags}
\end{figure}

We start by  constructing   the Euclidean correlator,
\begin{equation}\label{baa}
 \braket{T_{\mu\nu} (p) T_{\alpha\beta} (-p)}_{\mathrm{\E}}
 =  \braket{T_{\mu\nu}  T_{\alpha\beta} }_{\mathrm{\E}(0)}
 + \lambda \braket{T_{\mu\nu}  T_{\alpha\beta} }_{\mathrm{\E}(1)}
 +  \lambda^2 \braket{T_{\mu\nu} T_{\alpha\beta}}_{\mathrm{\E}(2)}
 + \mathcal O(\lambda^3)\,,
\end{equation} 
where the dependence on \(p\) in the rhs is understood.
The Feynman diagrams to order \(\mathcal O(\lambda^2)\) are shown in figure~\ref{fig:TTdiags}.
Except for the eye diagram \((2)f\), all integrals in the other diagrams' contributions can be treated with  two-propagator integral technology, summarized in appendix~\ref{app:formulae}.
The eye diagram is considerably more complicated, nonetheless it can be   computed exactly; details are in appendix~\ref{app:eye}. 
For the purpose of this paper,  we could disregard the terms with tensorial dependence on the external momentum, since they  vanish when contracted with the polarization tensor, \(\varepsilon^{\mu\nu}p_\nu=0\) when \( p^\mu=(q,\vec 0) \).
However, we provide the result for the \textit{full} Euclidean 2-point function with generic momentum  in   appendix~\ref{app:full2pt} as an additional technical result.

Next, we need to rotate from Euclidean to Lorentzian signature and construct the Wightman function. The Euclidean diagrams   are all  proportional to \((p^2)^{-\alpha}\), with the exponent fixed by dimensionality.
The relevant Wightman function follows from the prescription
\begin{equation}\label{acd}
\braket{T_{\mu\nu} (x) T_{\alpha\beta } (0)}  = \lim_{\epsilon \to 0^+}    \braket{T_{\mu\nu} (x_\mathrm{\E}) T_{\alpha\beta} (0)}_{\mathrm{\E}} \,,\qquad x^{\0}_\mathrm{\E} = i x^{\0} +\epsilon.
 \end{equation}
Constructing the Wightman function in momentum space is less immediate than in position space, and  requires straightforward  but tedious mathematical manipulations. 
In the case of the 2-point function,  the prescription \eqref{acd} translates  into the replacement \((p^2)^{-\alpha}  \to 2 \sin(\pi \alpha) \ \Theta[ p^{\0}-|\vec p\,|]\, |p^2|^{-\alpha} \), where on the rhs the Lorentzian metric is used. The step function selects   timelike momenta with positive energy, consistent with our choice \( p^\mu=(q,\vec 0), \,q>0 \).
We collect   more details of the Wick rotation in momentum space in appendix~\ref{app:WR}.

\begin{figure}[h]
	\centering
	\includegraphics[width=9cm]{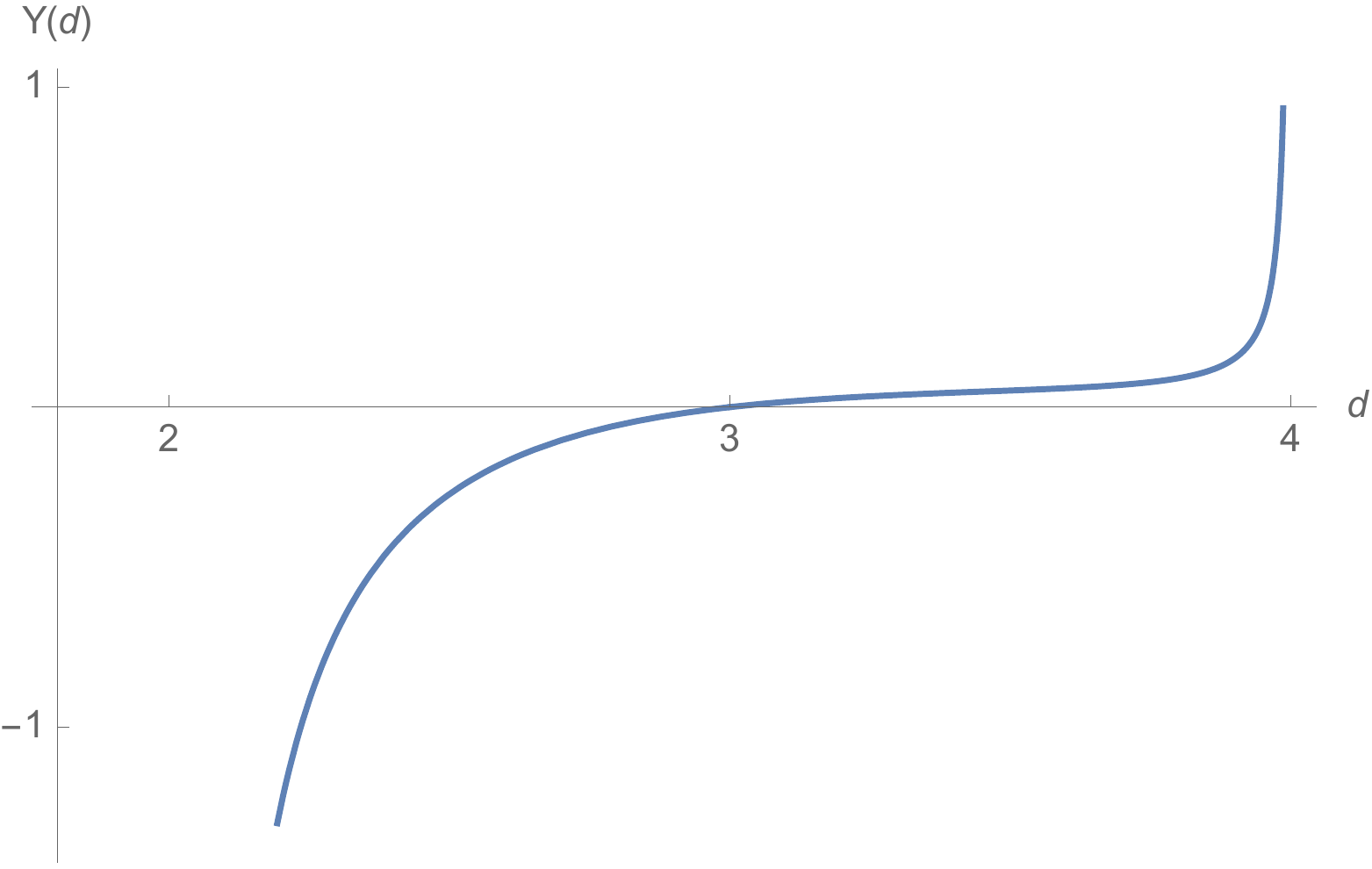}
	\caption{\(\Upsilon(d)\)  is negative  for  \(  2<d<3 \) and positive for \( 3 <d <4\), vanishes for \(d=3\) and has poles for \(d=2\) and \(d=4\),  increasing monotonically from \(\Upsilon(2^+)=-\infty\) to  \(\Upsilon(4^-)=+\infty\)}\label{Upsilon}
\end{figure}

Finally, the norm exhibits the form 
\begin{equation}\label{zba} 
      N =  a \      \tilde   \varepsilon_{ij}^* \tilde  \varepsilon_{ij} + b  \ 	 \varepsilon^*_{ii}\varepsilon_{jj} \,,
\end{equation}
where \( \tilde  \varepsilon_{ij} \) and \(\varepsilon_{ii}\) are the symmetric traceless part and the trace of the spacelike polarization tensor \(\varepsilon_{ij } = \varepsilon_{kk}\frac{\delta_{ij}}{d-1} + \tilde \varepsilon_{ij} \), and for the coefficients we find
{\small
\begin{align}\label{zaa}
a& = 
      		 \frac{  q^{d}  
		            }{  2^{d+3}\, (4\pi)^{\frac{d}2-\frac32} \,   \Gamma[\frac32 + \frac{d}2]  }
		     +\frac{\lambda^2  q^{3d-8}}{(4 \pi )^{\frac{3 d}{2}-1} }
		     \Bigg\{  \frac{(2 d-3) (324 - 434 d + 173 d^2 - 21 d^3)\,\Gamma [\frac{d}{2}-1] ^4 }{6 \,(d-4) (d-3) (d+1) \,\Gamma [\frac{3 d}{2}-3]   \,\Gamma[2 d-1]} 
		     \\
		     &
		     		    \hspace{60mm}
		     		    - \frac{ 4 \pi \,(d-6) (d-2)  \Gamma [3-\frac{d}{2}] \,\Gamma[3-\frac{3 d}{2}] \, \Gamma[  \frac{d}{2}-1 ]^3}{4^3 \, (d-3) (d+1)\,\Gamma [d]\, \Gamma [ \frac{3 d}{2}+\frac{1}{2}] \,\Gamma [\frac{1}{2}-\frac{3 d}{2}]} \nn
		    \\ &
		   \hspace{41mm}
		    +\frac{ (4\pi)^4
		     		(d-1)(2 \cos (\pi  d)+1) \csc (\pi  d) \,\Gamma [3-\frac{3 d}{2}]  \,\Gamma [ \frac{d}{2}-1]^2 
		     	}{
		     		4^{\frac{d}2+1}\, \,\Gamma  [\frac32 + \frac{d}2]  \,\Gamma [2 d-3]\, \Gamma[3-d] \,\Gamma[1-\frac{d}{2}]
		     	}  \left[  _3 {F}_2\left(\left.
		      \begin{smallmatrix}
		      1\; , \; 2-d\; , \; d-2
		      \\
		      3-d\; , \; 1-\frac{d}{2}
		      \end{smallmatrix}
		      \right\rvert 1 \right)\right. \nn\\
		     &
		     		    \hspace{32mm}  \left.
		     +\frac{3  (2 d-5) }{(d-3)} 
		     \, _3 {F}_2\left(\left.
		     \begin{smallmatrix}
		     1\; , \; 3-d\; , \; d-2 \\ 
		     4-d\; , \; 2-\frac{d}{2}
		     \end{smallmatrix}
		     \right\rvert 1 \right)
		     +
		     \frac{3 (d-2) (2 d-5) (3 d-8) }{2(d-4)^2(d-1)} 
		     \, _3 {F}_2\left(\left.
		     \begin{smallmatrix}
		     1 \; , \; 4-d \; , \; d-2\\
		     5-d \; , \; 3-\frac{d}{2}
		     \end{smallmatrix}
		     \right\rvert 1 \right)\right] 
		                     \Bigg\},\nn
\end{align}
\begin{align}     \label{zab}                       	             
b      &= 
         	 \frac{  q^{d} \  (d^2-1)    \left( \xi - \xi_d\right)^2  }{ 2^{d}\,   (4\pi)^{\frac{d}2-\frac32}\,   \Gamma[\frac32 + \frac{d}2]  }
             \left(
 	                   1 -
 	                   \frac{\lambda  }{q^{4-d}} \Upsilon(d)
                \right) 
                \\
                & \quad+\frac{\lambda^2  q^{3d-8}}{(4 \pi )^{\frac{3 d}{2}-1}}
                \Bigg\{\frac{\Gamma [\frac{d}{2}-1]^4}{ \Gamma [2 d-5]\, \Gamma[\frac{3 d}{2}-3]}
                \Bigg[\frac{2 \left(\xi - \xi_d \right)^2}{ (d-4)}\left(\frac{1}{3}+\frac{ (3 d-8) }{(d-4)}\, _3F_2\left(
                \left.
                \begin{smallmatrix}
                1 \; , \; 4-d \; , \; d-2\\
                5-d \; , \; 3-\frac{d}{2}
                \end{smallmatrix}
                \right\rvert 1
                \right)\right) \nn\\ 
                &\hspace{70mm} +\frac{(d-4) }{3 (d-3) (d-1)}\,\xi -\frac{(d-4) (7 d^2-29d+28)}{48 (d-3) (d-1)^2 (2 d-5)}
                \Bigg] \nn\\ 
                &\hspace{20mm}+{\left(\xi - \xi_d \right)^2 }\left[\frac{\Gamma [\frac{d}{2}-1]^6 \,\Gamma [2-\frac{d}{2}]^3}{4 \,\Gamma [d-2]^3\, \Gamma \left(4-\frac{3 d}{2}\right) \Gamma [\frac{3 d}{2}-3]}
                +\frac{(3 d-8)}{(d-3) }\frac{\Gamma \left(\frac{d}{2}-1\right)^3 \Gamma [2-\frac{d}{2}]}{\Gamma [\frac{3 d}{2}-3]\, \Gamma [d-2]}\,2 \pi  \cot \left(\frac{3 \pi  d}{2}\right) 
                \right]
                \Bigg\},\nn
\end{align}
}%
with  \(\Upsilon(d)\)   defined  as
\begin{equation}\label{cfh}
\Upsilon(d) =  
\frac{
	(d-1)\	\Gamma[3-\frac{d}2] \ \Gamma[\frac{d}2-1]
}{2^{d}\,
	(4 \pi )^{\frac{d}2-\frac32}\, (4-d)	\, \Gamma[\frac12+\frac{d}2]^2\ \Gamma[\frac12-\frac{d}2]
} 
\,.
\end{equation}   
The behaviour of this function is plotted in figure \ref{Upsilon}.

In \(d=4\) a finite result is obtained by adding the known  renormalization  counterterms\footnote{Explicitly we write \(\xi_\mathrm B=\xi+\delta\xi\,,\quad
	\mu^{d-4}\,\lambda_\mathrm B=\lambda+\delta\lambda\,, \quad
	\mu^{2-d/2}\,\phi_\mathrm B=\sqrt{1+\delta Z}\,\phi\), where the counterterms to the relevant order read
	\begin{equation}
	\begin{gathered}
	\delta\xi=\frac{\lambda(\xi-\frac16)}{(4\pi)^2(4-d)}+\frac{2\lambda^2(\xi-\frac16)}{(4\pi)^4(4-d)^2}-\frac{5\lambda^2(\xi-\frac{7}{30})}{12\,(4\pi)^4(4-d)} \,,\qquad
	\delta\lambda =\frac{3\lambda^2}{(4\pi)^2(4-d)}\,,
	\qquad
	\delta Z =-\frac{\lambda^2}{12(4\pi)^4(4-d)} \,.
	\end{gathered}
	\end{equation}
	\label{footnote-counterterms}
}  \cite{Toms:1982af}, which is a nontrivial consistency check of our general expression for the Euclidean 2-point function, and in particular of the contribution coming from the eye diagram.
After renormalization with minimal subtraction, we obtain
\begin{equation}\label{zak}
\begin{aligned}
a & =  \frac{ q ^4}{120\, (4\pi)}     \Big[   1 - \frac5{36} \frac{\lambda^2}{(4\pi)^4}    \Big]
\,,
\\
b& =  \frac{  q^4    }{  4\pi}
             \Big[ 
 	                     ( \xi - \tfrac16  )^2  +
 	                   \frac{\lambda  }{(4\pi)^2}  ( \xi - \tfrac16  )^2 
 	               \,     \Big( \log \frac{q^2}{\mu^2} -2 - \frac{1}{3(6\xi-1)}  \Big) 
 	                         + 
 	                         \frac{\lambda^2  }{864(4\pi)^4} \big[ -10 \pi ^2 (1-6 \xi )^2 
	  \\
  &\hspace{1.6cm} {}
+147+ 4 \xi  (1467 \xi -464) +30 (1-6 \xi )^2 \log^2 \frac{ q^2}{\mu^2} -12 (6 \xi -1) (70 \xi -11) \log      \frac{q^2}{\mu^2}
\big] \Big],
\end{aligned}	
\end{equation}
where  the renormalization scale  has been redefined as \(\mu^2 \to    \mu^2 \, e^{\gamma_E } /4 \pi\,   \).

The coefficient \(a\) does not receive \( \mathcal O(\lambda)\) corrections and is finite at order \( \mathcal O(\lambda^2)\). The coefficient \(b\) at first order in \(\lambda\) is  zero at the conformal value \(\xi=\xi_4=\frac16\); however, starting at second order in the coupling, the corrections are nonvanishing, corresponding to the quantum breakdown of classical conformal symmetry \cite{Toms:1982af, Callan:1970ze, Freedman:1974ze}.

In a CFT,  the term \(b\,|\varepsilon_{ii}|^2\) is absent from the norm, and the coefficient \(a\) of $|\tilde\varepsilon_{ij}|^2$   is related to the type-B trace anomaly coefficient $\mathrm{c}$, as\footnote{
The Euclidean 2-point function of the stress tensor has the form
\begin{align}
	{\varepsilon^*}^{\mu\nu}\,\varepsilon^{\alpha\beta}\,\langle T_{\mu\nu}(q)\,T_{\alpha\beta}(-q)\rangle_{{\mathrm E}}=
	\frac{q^4}{720\pi}\,\mathrm{c}\,\left(\frac{2}{ 4-d}-\log (\frac{q^2}{\mu^2})+ \mathcal O(4-d)
	\right)\,|\tilde\varepsilon_{ij}|^2\,,
\end{align}
where $\mathrm c$ is the trace anomaly
$180(4\pi)^2 \langle T\rangle=- \mathrm{a}\, \mathbb E_4+\mathrm{c}\, \mathrm{Weyl}^2$. 
After renormalizing   the UV divergence  by addition of a gravitational counterterm, the Euclidean correlator is finite and the anomaly $\mathrm{c}$ is the coefficient of the   logarithm. 
When Wick rotating to Lorentzian signature to construct the Wightman 2-point function, the Euclidean correlator gets multiplied by a factor of $(\Gamma[2-d/2] \,\Gamma[d/2-1])^{-1}=(4-d)/2+  \mathcal O(4-d)^2$, which renders the Euclidean pole finite and the logarithm disappears, ${\varepsilon^*}^{\mu\nu}\,\varepsilon^{\alpha\beta}\,\langle T_{\mu\nu}(q)\,T_{kl}(-q)\rangle=
\frac{q^4}{720\,\pi}\,\mathrm{c}\,|\tilde\epsilon_{ij}|^2.$  
}
$
a= \frac{ q ^4}{180 (4\pi)} \mathrm{c},
$
with $\mathrm{c}=3/2$ for the conformally-coupled free scalar.
This suggests a generalization of this  anomaly coefficient along the  RG flow,  as  
\begin{equation}
\mathrm{c} = \frac 32\,\left(1 - \frac5{36}\, \frac{\lambda^2}{(4\pi )^4}\right),
\end{equation}
matching the independent result of \cite{Hathrell:1981zb}. It would be interesting   to extend the calculation to higher orders and see  the logarithmic dependence on the energy scale.

\section{ANEC expectation value on the stress-tensor state}\label{sec-numerator}

In this section we present the evaluation of the correlator of the energy flux operator in terms of the Wightman 3-point function of stress-tensors,\footnote{For an alternative method to compute expectation values of the ANEC (and more generically of detector operators), based on using Feynman rules within the in-in formalism, see \cite{Caron-Huot:2022eqs}.}
\begin{equation}\label{tab}
\begin{aligned}
 \braket{\mathcal E }  &   = \lim_{z^+\to \infty} \Big(\frac {z^+}2 \Big)^{d-2} \intl_{-\infty}^{+\infty } \ddd{z^-} \int \ddd{^dx} e^{- iqx} {\varepsilon^*}^{\mu\nu}\varepsilon^{\alpha\beta}
 \braket{T_{\mu\nu}(x) \, T_{--}(z^{\pm}) \,T_{\alpha\beta}(0) }
 \\
 & =
 2\,  {\varepsilon^*}^{\mu\nu}\varepsilon^{\alpha\beta} \,
  \lim_{z^+\to \infty} \Big(\frac {z^+}2 \Big)^{d-2}\,
 \int\frac{d^{\scriptscriptstyle{d-1}}\vec{p}}{(2\pi)^{\scriptscriptstyle{d-1}}}\,\,\,e^{ 2i  p^{\1} r}\,
 \langle T_{\mu\nu}( q,\vec{0}\,)\,T_{--}(-p^{\1},\vec{p}\,)\,T_{\alpha\beta}(p^{\1}- q,-\vec{p}\,)\rangle\,.	
 \\
\end{aligned}
\end{equation} 
Following \cite{Hofman:2008ar}, for simplicity we inserted the ANEC operator at $(z^+\rightarrow\infty, z^-, z^a=0)$,
where \(z^{\scriptscriptstyle{\pm}} = z^{\1} \pm z^{\0}\), and \(z^a\) indicates the transverse directions \(a = 2,\ldots,d\).

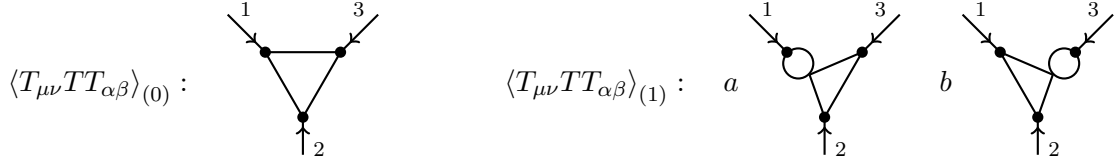
\begin{figure}
\centering
\label{fig:TTTdiags}
\( \braket{T_{\mu\nu}  T  T_{\alpha\beta} }_{ (0)} :\quad \)
 \begin{tikzpicture}[baseline={([yshift=-.5ex]current bounding box.center)},thick]
\draw[particle] (-1,0.5)--(-0.5,0);
\draw[particle] (1,0.5)--(0.5,0);
\draw[particle] (0,-1.73/2-0.5)--(0,-1.73/2);
\draw (-1/2,0) -- (1/2,0);
\draw (-1/2,0) -- (0,-1.73/2);
\draw (1/2,0) -- (0,-1.73/2);
\node at (-0.75,0.5){\({}^ 1\)};
\node at (0.75,0.5){\({}^ 3\)};
\node at (0,-1.73/2-0.5)[right]{\({}^ 2\)};
\node at (-0.5,0)[circle,fill,inner sep = 1.5pt] {};
\node at (0.5,0)[circle,fill,inner sep = 1.5pt] {};
\node at (0,-1.73/2)[circle,fill,inner sep = 1.5pt] {};
\end{tikzpicture} 
 \(\qquad\qquad \braket{T_{\mu\nu} T  T_{\alpha\beta} }_{ (1)} :\quad \)
              \(a\)
 \begin{tikzpicture}[baseline={([yshift=-.5ex]current bounding box.center)},thick]
\draw[particle] (-1,0.5)--(-0.5,0);
\draw[particle] (1,0.5)--(0.5,0);
\draw[particle] (0,-1.73/2-0.5)--(0,-1.73/2);
\draw (-0.2,-0.3) -- (1/2,0);
\draw (-0.2,-0.3) -- (0,-1.73/2);
\draw (1/2,0) -- (0,-1.73/2);
\node at (-0.5,0)[circle,fill,inner sep = 1.5pt] {};
\node at (0.5,0)[circle,fill,inner sep = 1.5pt] {};
\node at (0,-1.73/2)[circle,fill,inner sep = 1.5pt] {};
\draw (-0.35,-0.15) circle (0.2); 
\node at (-0.75,0.5){\({}^ 1\)};
\node at (0.75,0.5){\({}^ 3\)};
\node at (0,-1.73/2-0.5)[right]{\({}^ 2\)};
\end{tikzpicture} 
\quad
             \(b\)
 \begin{tikzpicture}[baseline={([yshift=-.5ex]current bounding box.center)},thick]
\draw[particle] (-1,0.5)--(-0.5,0);
\draw[particle] (1,0.5)--(0.5,0);
\draw[particle] (0,-1.73/2-0.5)--(0,-1.73/2);
\draw (0.2,-0.3) -- (-1/2,0);
\draw (0.2,-0.3) -- (-0,-1.73/2);
\draw (-1/2,0) -- (0,-1.73/2);
\node at (-0.5,0)[circle,fill,inner sep = 1.5pt] {};
\node at (0.5,0)[circle,fill,inner sep = 1.5pt] {};
\node at (0,-1.73/2)[circle,fill,inner sep = 1.5pt] {};
\draw (0.35,-0.15) circle (0.2); 
\node at (-0.75,0.5){\({}^ 1\)};
\node at (0.75,0.5){\({}^ 3\)};
\node at (0,-1.73/2-0.5)[right]{\({}^ 2\)};
\end{tikzpicture} 
\caption{Diagrams for the 3-point function. We did not include the third permutation of the first order diagram because it does not contribute to the expectation value.}
\label{fig:TTTdiags1}
\end{figure}

We construct \( \braket{\mathcal E } \) starting from   \(\braket{T_{\mu\nu} (x)\, T(z) \, T_{\alpha\beta}(y)}_\mathrm{\E} \) where 
\be
T \equiv T_{--} = \partial_-\phi\,\partial_-\phi 
\ee
 is the component of the stress tensor in the Euclidean null direction. We ignore \(\xi\) terms in this operator because they are total derivatives that vanish after the integration over \(z^-\).
We write the momentum-space perturbative expansion as
\begin{equation}\label{bba}
 \braket{T_{\mu\nu} (p_1) \, T (p_2) \, T_{\alpha\beta} (p_3)}_{\mathrm{\E}}
 =  \braket{T_{\mu\nu}T  T_{\alpha\beta} }_{\mathrm{\E}(0)}
 + \lambda \braket{T_{\mu\nu}  TT_{\alpha\beta} }_{\mathrm{\E}(1)} 
 + \mathcal O(\lambda^2).
\end{equation} 
Diagrams with a 2-propagator subdiagram depending only on the momentum \(p_2\) can also be discarded. This is justified because such integral is proportional to  \((p_{2\-})^2\), which vanishes upon integration over \(z^-\). To slightly simplify the expressions we identify  \( p_{2\-} =0=-p_{1\-} - p_{3\-} \) from the start. The Feynman diagrams up to order \(\mathcal O(\lambda^1)\) are in figure~\ref{fig:TTTdiags1}.
The Euclidean correlators read
\begin{equation}\label{3ptE} 
\begin{aligned}
\braket{T_{\mu\nu}T  T_{\alpha\beta} }_{ \mathrm E (0)}& = 
8 
\int \!\!    {\dbar^d k}    \
\frac{[{(k-p_1)}_{\-}]^2   \ V_{\mu\nu} (k,p_1) \ V_{\alpha\beta}(k,-p_3)}{k^2 \, (k+p_3)^2 \, (k-p_1)^2}  \,,
\\
\braket{T_{\mu\nu}T  T_{\alpha\beta} }_{ \mathrm E (1)a} 
&   = 
4 \frac{ (\xi-\xi_d) \Gamma[\frac{d}2]^2 \,\Gamma[1-\frac{d}2]}{ { (4\pi)^{\frac{d}2}}  \,\Gamma[d-1]}  
    \frac{ p_{1\mu}p_{1\nu}-p_1^2  \delta_{\mu\nu}  }{  [(p_1)^2]^{2-d/2}} \!
  \int\! \dbar^d k \,
        \frac{  [(k-p_1)_{\-}]^2\, W_{\mu \nu \alpha \beta}(k,p_3)}{k^2 \, (k-p_1)^2 \, (k+p_3)^2 } \,  ,   
\\
\braket{T_{\mu\nu}T  T_{\alpha\beta} }_{\mathrm E  (1)b} &  = \braket{T_{\alpha\beta}T  T_{\mu\nu} (p_1 \leftrightarrow  p_3)}_{\mathrm E { (1)a} }\,,
\end{aligned}	   
\end{equation}   
where 
\begin{equation}\label{VW}
\begin{aligned}
V_{\alpha\nu }  (k,p) &=
k_\alpha k_\nu - k_{  (\alpha }  p_{\nu ) }  + \xi\, p_\alpha p_\nu 
- \frac12 \delta_{\alpha\nu} ( k^2  - k\cdot p+2 \xi \,p^2 )  \,,\\
W_{\mu\nu\alpha\beta}(k,p ) &
= 
2 k_\alpha k_\beta -\delta_{\alpha  \beta } k^2 -\delta_{\alpha  \beta }\,k\cdot p+2 k_{ (\alpha}   p_{\beta ) }
+2 \xi(p_{\alpha} p_{\beta} -\delta_{\alpha\beta }{p}^2  )  \,.
\end{aligned}
\end{equation}

The Wightman function relevant to compute the expectation value is the one defined as
\begin{equation}\label{aia}
\braket{T_{\mu\nu} (x) \ T(z) \ T_{\alpha\beta } (0)}  = \lim_{\epsilon,\zeta \to 0^+}    \braket{T_{\mu\nu}  (x_\mathrm{\E})  T(z_\mathrm{\E}) T_{\alpha\beta} (0)}_{\mathrm{\E}} \,,\qquad
	x^{\0}_\mathrm{\E} = i x^{\0}+ \epsilon \,,\quad 
	z^{\0}_\mathrm{\E} = i z^{\0} + \zeta \,,
\end{equation}
with \(\epsilon > \zeta\). As opposed to the case of the 2-point function, where the momentum dependence is particularly simple, here the analyticity properties of the momentum-space correlators are more complicated and no simple prescription for the Wick rotation from Euclidean to Lorentzian signature is available.
One has to resort to a case-by-case analysis. We give further   details of the Wick rotations  needed for \eqref{3ptE} in appendix~\ref{app:WR}; a complete discussion of the method used can be found in \cite{Bautista:2019qxj,Casarin:2021fgd}.

The resulting Lorentzian expressions  read
\begin{equation}\label{aif} 
\begin{aligned}
\braket{T_{\mu\nu}T  T_{\alpha\beta} }_{ (0)}& = 
8\,(2\pi)^3 
\int \!\!    {\dbar^d k}    \ [{(k-p_1)}_{\-}]^2 \ \hat V_{\mu\nu} (k,p_1) \ \hat V_{\alpha\beta}(k,-p_3)   \ 
\
\bar \delta[p_1- k] \ \dot  \delta [p_3 + k ] \ \bar  \delta[k] \,,
\\
\braket{T_{\mu\nu}T  T_{\alpha\beta} }_{ (1)a} 
&   = 
 \frac{ (\xi-\xi_d) \Gamma[\frac{d}2]^2 \,\Gamma[1-\frac{d}2]}{ {2 (4\pi)^{\frac{d}2-3}}  \,\Gamma[d-1]}  
    \frac{ p_{1\mu}p_{1\nu}-p_1^2  \eta_{\mu\nu}  }{  |(p_1)^2|^{2-d/2}} \!
  \int\! \dbar^d k \,[(k-p_1)_{\-}]^2\,
      	    	             	          	          	    \hat W_{\mu \nu \alpha \beta}(k,p_3)\,    
      	    	       	          	  	  \dot  \delta [ p_3+k ] \times
      	    	     \\[-0.25em]
      	    	     & \qquad    \hspace{2mm}   \times
      	     \left[  \cos \big[\frac{ \pi d}{2} \big] \ \bar \delta[k] \ \dot \delta[k-p_1]
      	          	  -	  \frac{1	}{\pi}\sin \big[ \frac{ \pi d}{2} \big]\,\Theta[p_1^{\0}-|\vec p_1|]
      	    \left(
     	    \frac{ 
     	       \bar \delta[k]
  	        	          	  }{ (k-p_1)^2 }   +
     	        \frac{ 
     	             	       \bar \delta[k-p_1]
  	        	             	          	  }{k^2  }   
  	        	          	   \right)
      	        \right]\,,
\\
\braket{T_{\mu\nu}T  T_{\alpha\beta} }_{ (1)b} &  = \braket{T_{\alpha\beta}T  T_{\mu\nu} (p_1 \leftrightarrow -p_3)}_{ (1)a} \,,
\end{aligned}	   
\end{equation}   
with \(\hat  V\) and \(\hat W\) the same as \eqref{VW} but with the Lorentzian metric \(\delta \to \eta \),
and where we have used the notation
\begin{equation}
\bar\delta[k]\equiv\frac{\delta[k^{\0}-|\vec k|]}{k^{\0}+|\vec k|}\,,\qquad \dot\delta[k]\equiv\frac{\delta[k^{\0}+|\vec k|]}{-k^{\0}+|\vec k|}\,.
\end{equation}

These expressions can be directly used to evaluate the correlator of the energy flux \eqref{tab}. The delta function   \(  \dot  \delta [ p_3+k ] \) can be used to  integrate  \(p_3^1\).
The large-\(z^+\) limit can then be computed rescaling of the transverse components \(p_3^a \to p_3^a/z^+\), which makes manifest the falloff \(\sim (z^+)^{2-d}\) of the correlator, and the remaining integrals can be done relatively easily.\footnote{This type of calculations are described in detail in   \cite{Casarin:2021fgd}.}   

The energy flux operator \( \mathcal E\) breaks the \(\mathrm{SO}(d-1)\) rotational invariance to \( \mathrm{SO}(d-2) \) corresponding to the rotations in the transverse directions. We can therefore further decompose the polarizations as \(\varepsilon_{ij} \simeq (\varepsilon_{11} , \varepsilon_{1a} , \tilde \varepsilon_{ab}, \varepsilon_{aa})\) corresponding to the scalar component \(11\), a vector, a symmetric traceless tensor and the trace part in the transverse directions.
The general tensor decomposition of the correlator of the energy flux in terms of these polarizations is therefore
\begin{equation}
\begin{aligned}\label{qbd}
 \braket{ \mathcal{E}} 
        &= \hat a \    | \varepsilon_{aa} | ^2 
        +  \hat b\  | \varepsilon_{11} | ^2  
        + \hat c  \ \left(\varepsilon_{11} \varepsilon^*_{aa}  +   \varepsilon^*_{11} \varepsilon_{aa}   \right)
       +  \hat e \ \varepsilon^*_{a1}  \varepsilon_{a1} 
       +  \hat f \ \tilde  \varepsilon^*_{ac}\tilde  \varepsilon_{ac}\,.
\end{aligned}
\end{equation}
To order \(\mathcal O(\lambda^1)\) we obtain the following expressions for these coefficients:
\begin{equation}\label{cfc}
\begin{aligned}  
\hat a & =\frac{ 
 		 	 q^{d+1} \ (4 \xi - 1)^2
 		}{  
 		 16 	(4\pi)^{d-2}   
 		} 
 		\left(
 			1 -	4\frac{ (\xi - \xi_d) }{(4 \xi - 1 )} 
 			\frac{\lambda}{ q^{4-d}  } \Upsilon(d) 
 		\right)\,,
\quad   \hat b   = \frac{ 
 	 	 q^{d+1} \ \xi^2
 	}{  
 	 	(4\pi)^{d-2}   
 	}  
 		\left(
 			1 -
 			 	\frac{(\xi - \xi_d) }{\xi} 
 		\frac{\lambda}{ q^{4-d} } \Upsilon(d)
 		\right)\,,
\\ \hat c & = \frac{ 
 	 	 q^{d+1}  (4\xi-1) \xi
 	}{  
 	  4 	(4\pi)^{d-2}   
 	}  
 		\left(
 			1 -
 			\frac{	(\xi - \xi_d)  (8\xi-1)}{2 \xi (4 \xi - 1 )   } 
 			\frac{\lambda}{ q^{4-d} } \Upsilon(d)
 		\right)\,,
 		 \qquad\qquad
  \hat e =0=\hat f\,,
\end{aligned}
\end{equation} 
where \(\Upsilon(d)\) is the one defined  in \eqref{cfh}.  We note the relation \(\hat c^2 = \hat a \, \hat b  + \mathcal O(\lambda^2).\)

The  case \(d=4\) is special because renormalization is required. Adding the appropriate counterterms (cf.\ footnote \ref{footnote-counterterms}), the coefficients of the energy flux correlator become
\begin{equation}\label{cfj}
\begin{aligned}
\hat a&=
\frac{ 
 		 	 q^{5} \ (4 \xi - 1)^2
 		}{  
 		 16 	(4\pi)^{2}   
 		} 
 		\Big[ 
 			1  +\frac{\lambda}{(4\pi)^2} \frac{\xi-\tfrac16}{\xi-\tfrac14} 
 			\Big(
 			 \log \frac{ q ^2}{\mu^2} -2 -  \frac1{3(6\xi-1) }
 			\Big)
 		\Big]
\,, \\
\hat b&=
\frac{ 
 		 	 q^{5} \ \xi^2
 		}{  
 		 16 	(4\pi)^{2}   
 		} 
 		\Big[ 
 			1  +\frac{\lambda}{(4\pi)^2} \frac{\xi-\tfrac16}{\xi} 
 			\Big(
 			 \log  \frac{ q ^2}{\mu^2} -2 -  \frac1{3(6\xi-1) }
 			\Big)
 			\Big]
\,, \qquad
\hat c^2 = \hat a \, \hat b  + \mathcal O(\lambda^2)\,.
\end{aligned}
\end{equation}
In these expressions we observe that the values \(\xi = 0,1/4\) emerge naturally. We will see how these play a role below.

\section{Positivity bounds}\label{sec-bounds}

The ANEC demands the (normalized) expectation value,
\begin{equation}
\begin{aligned}\label{taa}
\braket E & = \frac{\braket{\mathcal E }}{N}\,,
\end{aligned}  
\end{equation}
to be non-negative. Since the norm of the state $N$  has to be positive by unitarity, the non-normalized correlator \(\braket{\mathcal E }\), given by the energy flux correlator, is also non-negative.

Given the norm  $
N =  a \        |\tilde \varepsilon_{ij}|^2 + b  \ 	 |\varepsilon_{ii}|^2$,
unitarity translates into 
\be\label{cug}
a\geq0,\qquad b\geq0,
\ee
 whose expressions are given in \eqref{zaa} and \eqref{zab} up to $\mathcal O(\lambda^2)$.  
 Given the complexity of the expressions, we will consider the unitarity constrain at full \(\mathcal{O}(\lambda^2)\) only in \(d=4\). In \(2<d<4\) we will only consider the expression at \(\mathcal{O}(\lambda^1)\), which is the order at which we evaluate the ANEC correlator.

The expectation value \(\braket {  E} \) is the energy flux for unit angle in the \(S_{d-2}\) sphere, and therefore   reduces to the energy \( q\) when integrated over it. The general expression can therefore be written as
\begin{equation}\label{qca}
\begin{gathered}
\braket{ \mathcal {E}} 
=
K\
\Big[   a \      \tilde   \varepsilon_{ij}^* \tilde  \varepsilon_{ij} + b  \ 	 \varepsilon^*_{ii}\varepsilon_{jj}+ 
c \left(
\varepsilon^*_{ii}\varepsilon_{11}+\varepsilon_{ii}\varepsilon^*_{11} + c_0
\right) +  
e \left( 
\varepsilon_{1i}^* \varepsilon_{1i}+ e_0
\right)+
f \left( 
\varepsilon_{11}^*  \varepsilon_{11}+ f_0
\right)
\Big]\,,
\\
K = \frac{ q}{\mathrm{Vol} {[S_{d-2}]}} 
\,, \qquad
c_0 = -\frac{2\,\varepsilon_{ii}^* \varepsilon_{jj} }{d-1} 
\,,
\qquad
e_0 =  -\frac{ \varepsilon_{ij}^* \varepsilon_{ij} }{d-1}
\,,
\qquad
f_0 =  - \frac{\varepsilon_{ii}^* \varepsilon_{jj} 
	+ 2 \varepsilon_{ij}^* \varepsilon_{ij}}{d^2-1 } \,,
\end{gathered}
\end{equation}
where \(a\) and \(b\) are the same as in the 2-point function \eqref{zba} and the constants \(c_0,e_0,f_0\) are chosen   to make each term in round brackets vanish when integrated over the \(S_{d-2}\) sphere.  The overall normalization \(K\) is chosen so that the first two terms have the same coefficients as in \eqref{zba}.  Via the decomposition  \(\varepsilon_{ij} \simeq (\varepsilon_{11} , \varepsilon_{1a} , \tilde \varepsilon_{ab}, \varepsilon_{aa})\)  we match \eqref{qca} with the decomposition \eqref{qbd}, to find
\begin{equation}\label{qcb}
\begin{gathered}
\frac{\hat a }{K}   =   \frac1 {(d-1)(d-2)}  a+b - \frac 2 {d-1} c - \frac 1 {(d-1)(d-2)} e - \frac d {d^2-1} f  \,,
\\
\frac{\hat b }{K}   =   \frac{d-2}{d-1}a+b +\frac {d-2} {d-1} (2c+e) + \frac {d^2-4} {d^2-1} f  \,,
\qquad\qquad
\frac{\hat c }{K}   = -\frac{1}{d-1}a+ b+ \frac {d-3} {d-1} c - \frac 1 {d^2-1}f \,,
\\
\frac{\hat e }{K}  =  2 a + \frac {d-3} {d-1} e - \frac 4 {d^2-1} f  \,,
\qquad\qquad
\frac{\hat f }{K}  =   a - \frac 1 {d-1} e - \frac 2 {d^2-1} f   \,.
\end{gathered}
\end{equation}
It is easier to express the ANEC positivity constraints for the hatted coefficients,
\begin{equation}\label{constr}
\hat a\geq0\,, \qquad
\hat b\geq0\,, \qquad
\hat a \hat b -  \hat c^2\geq0\,, \qquad
\hat e\geq0\,, \qquad
\hat f\geq0\,.
\end{equation}

In the particular case under consideration the situation is somewhat simpler.
The coefficients \eqref{cfc} of the tensor decomposition happen to satisfy  the relation
\(\hat a\, \hat b = \hat c^2  +\mathcal O(\lambda^2) \). This allows us to write \eqref{qbd} as the manifestly-positive expression
\begin{equation}
\begin{aligned}\label{cff}
 \braket{ \mathcal{E}} 
        &=   \left|\sqrt{\hat a} \  \varepsilon_{aa}  + \sqrt{\hat b} \ \varepsilon_{11}  \right| ^2 
\end{aligned}
\end{equation}
\emph{provided} \( \hat a \geq 0 \) and \(\hat b \geq 0\). Any other sign configuration would result in manifestly non-positive expressions, which are ruled out by the ANEC. Therefore,  three of the five constraints \eqref{constr}  are saturated.

We next spell out the constraints   explicitly depending on the dimension.
We write $\lambda\to \lambda\,\mu^{4-d}$, where $\mu$ is some energy scale (which in \(d=4\) becomes the renormalization scale) and \(  \lambda\) is  now dimensionless.

We have computed the relevant 3-point correlators at first order. At this order,  a violation of the ANEC and unitarity inequalities  corresponds to a breakdown of perturbation theory. 
  To proceed, we make the assumption that there is a $\xi$-independent range of  $\lambda$ small enough where the first-order approximation is valid. To assess the reliability of the bounds derived, we would need to go to the next order. The calculations involved for this lie at the boundary of current diagrammatic technology, and we leave this for future work.

\subsection[Case \(  {2<d<4}\).]{Case \(\boldsymbol {2<d<4}\).}

The ANEC constraints are  
\begin{equation}\label{cfg-2} 
\frac{\xi - \xi_d }{\xi-\frac14} \,
  \lambda \, \Upsilon(d) \left( \frac{\mu }{ q } \right)^{4-d}
\leq 1
\,,\qquad\qquad
\frac{\xi - \xi_d }{\xi} 
\,
  \lambda \, \Upsilon(d) \left( \frac{\mu }{ q } \right)^{4-d}
\leq 1
\,,
\end{equation}
at first order in perturbation theory.

Notice that if \(\xi=\xi_d\) the inequalities \eqref{cfg-2} are trivially satisfied. We therefore assume \(\xi \neq \xi_d\). 
From the left-hand side of the inequalities \eqref{cfg-2}, it emerges that the values $\xi=0$ and $\xi=\frac14$  play a distinguished role. We will see below that indeed these values flag forbidden ranges of the parameter. One explanation for the origin of these values might be that for \(d>2\), the conformal value $\xi_d=\frac{(d-2)}{4(d-1)}$ ranges from $\xi_{d=2}=0$ to $\xi_{d\rightarrow\infty}=\frac14$. It would be good to better understand this curious emergence.

To compare with the constraints from unitarity  at this order, the latter reduce to only one constraint, since the coefficient $a$ in \eqref{zaa} has no order $\mathcal O(\lambda^1)$ correction and the tree-level contribution is manifestly positive. We get
\begin{equation}\label{norm}
  \lambda \, \Upsilon(d) \left( \frac{\mu }{ q } \right)^{4-d}\leq 1\,. 
\end{equation}
 This constraint does not depend on $\xi$.

We now set the arbitrary scale to \(\mu=q\) (which sets the normalization of the dimensionless $\lambda$) and derive the consequences for the couplings at this scale. We find it convenient to further distinguish the cases of \(d\) smaller and bigger than \(d=3\),  which correspond to opposite signs for \(\Upsilon(d)\). At $d=3$, these bounds don't constrain $\xi$ or $\lambda$ since $\Upsilon(3)=0$.

\subparagraph{\(\boldsymbol{3<d<4}\).}  

In this case $\Upsilon(d)>0$ (see figure \ref{Upsilon}), and $\frac18<\xi_d<\frac16$, both increasing monotonically with $d$. The unitarity constraint \eqref{norm}, 
\begin{equation}\label{nor}
0<  \lambda \, \leq \frac{1}{ \Upsilon(d) } \,,
\end{equation}
puts an upper bound for the coupling \(\lambda\), which disappears as $d$ approaches $d=3$.

\begin{figure}
\centering
\begin{tikzpicture}[decoration={markings,
	mark=at position 0.58 with {\arrow[line width=1pt]{>}},
}
]
\draw[help lines,->] (-6,0) -- (9.5,0) coordinate (xaxis);
\node at (10,0) {$\xi$};
\node at (0,-0.5) {$0$};
\node at (3,-0.5) {$\frac14$};
\node at (-2,-0.5) {$-\Upsilon(d)\,\xi_d\,\lambda$};
\node at (6.1,-0.5) {$\frac14+\Upsilon(d)\,\left(\frac14-\xi_d\right)\,\lambda$};
\draw[color=red,line width=1pt,decorate,decoration={zigzag,amplitude=2,segment length=8,post length=0}] (-2,-0.001) -- (0,0);
\draw[color=red,line width=1pt,decorate,decoration={zigzag,amplitude=2,segment length=8,post length=0}] (3.001,0) -- (4.5,0);
\draw[color=black,line width=1pt,<-] (-5.5,0.01) -- (-2,0.01);
\draw[color=black,line width=1pt] (0,0.01) -- (3,0.01);
\draw[color=black,line width=1pt,->] (4.5,0.01) -- (8.5,0.01);
\draw[color=black,line width=1pt] (0,-0.15) -- (0,0.15);
\draw[color=black,line width=1pt] (-2,-0.15) -- (-2,0.15);
\draw[color=black,line width=1pt] (3,-0.15) -- (3,0.15);
\draw[color=black,line width=1pt] (4.5,-0.15) -- (4.5,0.15);
	\end{tikzpicture} 
	\caption{The ANEC bounds exclude two ranges of $\xi$, in red zig-zag lines, below $\xi=0$ and above $\xi=1/4$ when $3<d<4$.}
	\label{ranges xi}
\end{figure}
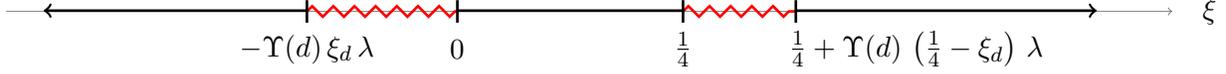

 This bound combined with the ANEC constraints limit the allowed range of $\xi$ at fixed $\lambda$ (recall that $0<\xi_d<1/4$):
\begin{equation}\label{3d4} 
\begin{aligned}
\xi \leq
- \Upsilon(d)\, \xi_d \,\lambda +\mathcal O(\lambda^2),
\quad\quad  \quad\quad 
0<\xi <\frac{1}{4},
 \quad\quad  \quad\quad 
   \xi \geq
   \frac{1}{4}+\Upsilon{(d)}  \,\left(\frac14-\xi_d\right) \,\lambda  +\mathcal O(\lambda^2),
\end{aligned} 
\end{equation}
where the $\lambda$-dependent upper and lower bounds have been expanded to first order in $\lambda$. Two gaps in the allowed range of $\xi$ appear around $\xi=0,1/4$, see figure \ref{ranges xi}. 
The gaps increase with the dimension $d$.
At exactly $d=4$, the bounds have to be analized separately because of renormalization.

Turning it around, we can derive bounds for $\lambda$ at fixed $\xi$. For $0<\xi<\frac14$, the unitarity bound \eqref{nor} is the most stringent, while for other values of $\xi$, the ANEC constraints 
 win:
 \be\label{brut}
 \xi<0\,:\quad\lambda\leq\frac{1}{ \Upsilon(d) } \frac{\xi}{\xi-\xi_d},\qquad\qquad \xi>\frac14\,:\quad\lambda\leq\frac{1}{ \Upsilon(d) } \frac{\xi-\frac14}{\xi-\xi_d}.
 \ee
 Again, these bounds become more stringent at $d$ close to $d=4$, where the upper bounds are small. As $d$ approaches $d=3$, they become automatically satisfied.

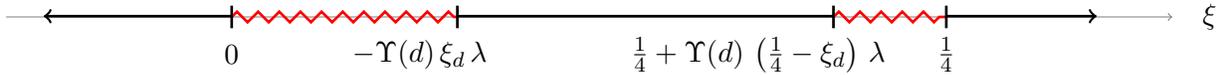
\begin{figure}[H]
	\centering
	\begin{tikzpicture}[decoration={markings,
		mark=at position 0.58 with {\arrow[line width=1pt]{>}},
	}
	]
	\draw[help lines,->] (-6,0) -- (9.5,0) coordinate (xaxis);
	\node at (10,0) {$\xi$};
	\node at (-0.5,-0.5) {$-\Upsilon(d)\,\xi_d\,\lambda$};
	\node at (4,-0.5) {$\frac14+\Upsilon(d)\,\left(\frac14-\xi_d\right)\,\lambda$};
	\node at (-3,-0.5) {$0$};
	\node at (6.5,-0.5) {$\frac14$};
	\draw[color=red,line width=1pt,decorate,decoration={zigzag,amplitude=2,segment length=8,post length=0}] (-3,-0.001) -- (0,0);
	\draw[color=red,line width=1pt,decorate,decoration={zigzag,amplitude=2,segment length=8,post length=0}] (5.001,0) -- (6.5,0);
	\draw[color=black,line width=1pt,<-] (-5.5,0.01) -- (-3,0.01);
	\draw[color=black,line width=1pt] (0,0.01) -- (5,0.01);
	\draw[color=black,line width=1pt,->] (6.5,0.01) -- (8.5,0.01);
	\draw[color=black,line width=1pt] (0,-0.15) -- (0,0.15);
	\draw[color=black,line width=1pt] (-3,-0.15) -- (-3,0.15);
	\draw[color=black,line width=1pt] (5,-0.15) -- (5,0.15);
	\draw[color=black,line width=1pt] (6.5,-0.15) -- (6.5,0.15);
	\end{tikzpicture} 
	\caption{The ANEC bounds exclude two ranges of $\xi$, in red zig-zag lines, abovew $\xi=0$ and below $\xi=1/4$ for $2<d<3$.}
	\label{ranges xi 2d}
\end{figure}

\subparagraph{\(\boldsymbol{2<d<3}\).} 
In this case $\Upsilon(d)<0$, and $0<\xi_d<\frac18$. The unitarity bound \eqref{norm} is automatically satisfied for any positive \(  \lambda\), in contrast to the previous case. 
The allowed regions for $\xi$ are now  entirely due to the ANEC and read  
\begin{equation}\label{2d3}
\begin{aligned}
\xi<0,
 \quad \quad \quad
-  \Upsilon(d)\,\,\xi_d \,\,\lambda +\mathcal O(\lambda^2)\,\,
\leq\,\,  \xi  \,\,\leq \,\,
\frac14+\Upsilon{(d)}  \,\left(\frac14-\xi_d\right) \,\lambda+\mathcal O(\lambda^2)  ,
 \quad \quad\quad
 \xi > \frac14 \,,
\end{aligned}
\end{equation}
where again the $\lambda$-dependent upper and lower bounds have been expanded to first order.
In writing \eqref{2d3} we  made  the additional simplifying assumption $\Upsilon(d)\,\lambda>-1$ to expand to first order, though it is not required by the constraints. 
The bounds \eqref{2d3} are plotted in figure \ref{ranges xi 2d}.

Alternatively, for certain values of $\xi$, we obtain upper bounds for  $\lambda$,
 \be
0<\xi<\xi_d\,:\quad 0< \lambda\leq\frac{1}{ \Upsilon(d) } \frac{\xi}{\xi-\xi_d},\qquad\qquad \xi_d<\xi<\frac14\,:\quad 0< \lambda\leq\frac{1}{ \Upsilon(d) } \frac{\xi-\frac14}{\xi-\xi_d}.
\ee
 Again, these bounds become trivial as $d$ approaches $d=3$, where the upper bounds diverge.

\subsection[{Case \(  {d=4}\).}]{Case \(\boldsymbol {d=4}\).}
Given the structure of \eqref{cfj} and \eqref{zak}, even though the constraints  \eqref{cfg-2} and \eqref{norm} diverge if we simply evaluate them at $d=4$ (because of $\Upsilon(d)$), they are still formally valid under the replacements \(\xi_d \to \frac 16\) and \(\Upsilon(d) \,  (\frac \mu q)^{4-d} \to  \frac{1}{(4\pi )^2 }  \big[  2 +  \frac1{3(6\xi-1) }-\log  \frac{ q ^2}{\mu^2}  \big] \). Furthermore, we introduce \(\hat \lambda = \lambda / (4\pi)^2\), which emerges as natural perturbative parameter. The ANEC constraints read
\begin{gather}\label{cfa}
\hat{\lambda} \ \frac{\xi-\tfrac16}{\xi-\tfrac14} 
\Big(
2 + \frac1{3(6\xi-1) }-\log  \frac{ q ^2}{\mu^2}
\Big)
\leq 1
, \qquad
\hat{\lambda} \ \frac{\xi-\tfrac16}{\xi} 
\Big(
2 + \frac1{3(6\xi-1) }-\log \frac{ q ^2}{\mu^2}
\Big)
\leq 1,
\end{gather}
and the unitarity one, at first order,
\begin{gather}\label{a}
	\hat{\lambda} \ 
	\Big(
	2 + \frac1{3(6\xi-1) }-\log  \frac{ q ^2}{\mu^2}
	\Big)
	\leq 1
.
\end{gather}
Looking back at the $\hat a$ and $\hat b$  coefficients \eqref{cfj} from which \eqref{cfa} follow, the first one is multiplied by a factor of $(\xi-\frac14)^2$, and the second one is multiplied by a factor of $\xi^2$. Therefore, the ANEC constraints are saturated and automatically satisfied in these cases. 
Similarly, the unitarity constraint \eqref{a} appears in fact multiplied by a factor of $(\xi-\frac16)^2$ from the  coefficient $b$ in \eqref{zak}. Therefore, when \(\xi=\frac16\), both in the free (conformal) $\lambda=0$ case  and in the interacting  case at first order, the unitarity inequality is saturated and trivially satisfied, and the pole in \eqref{a} is naturally avoided. 
 
Setting the   renormalization scale to \(\mu=q\), the logarithms disappear. The couplings \(\lambda\) and \(\xi\) have   an implicit dependence  on the scale due to renormalization; the inequalities become therefore conditions for the couplings evaluated at such value of the energy scale. However,   the running of \(\lambda\) starts at   \(\mathcal O (\lambda^2)\), and the running of \(\xi\) starts at    \(\mathcal O (\lambda^1)\) but it always appears multiplied by \(\lambda\). Therefore, since we work at first order in \(\lambda\), our inequalities  \eqref{cfa} do not capture the implicit dependence on \(\mu\). 

\paragraph{Unitarity bounds.}
Since we have computed the ANEC correlator at \(\mathcal O (\lambda)\), we first consider   the unitarity constraints at the same perturbative order. A big difference with the $2<d<4$ case, in the $d=4$ case the unitarity constraint depends on $\xi$. This allows in principle to set bounds on $\xi$ for fixed $\hat\lambda$ from this condition.

Using \eqref{zak}, \(a>0\) is automatically satisfied at first order, and \(b>0\) gives
\begin{equation} \label{uni1}
	\begin{aligned}  
\xi \leq \frac{1}{6}, \  \qquad\qquad\  
		\xi\,\, \geq\,\,\frac16+\frac{\hat\lambda}{18}+\mathcal O(\hat\lambda^2).
	\end{aligned}
\end{equation}  
This first order analysis seems to indicate that, at fixed $\hat\lambda$, a finite range of \(\xi\) above $\xi=\frac16$  is forbidden by unitarity. 
 
Extending the analysis to \(\mathcal O(\lambda^2) \), there is now also the condition from $a\geq0$. Putting both unitarity conditions $a,b\geq0$ together, the result that we find is that $\xi$ is no longer constrained. This is not a contradiction because the first-order bounds arise from the boundary of validity of perturbation theory. This is however a clear reminder that any of the bounds we derive need to be analysed in view of the higher order contributions, which   is outside the scope of this paper.

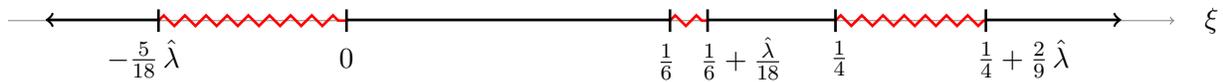
\begin{figure}[H]
	\centering
	\begin{tikzpicture}[decoration={markings,
		mark=at position 0.58 with {\arrow[line width=1pt]{>}},
	}
	]
	\draw[help lines,->] (-6,0) -- (9.5,0) coordinate (xaxis);
	\node at (10,0) {$\xi$};
	\node at (-1.5,-0.5) {$0$};
	\node at (5.05,-0.5) {$\frac14$};
	\node at (2.75,-0.55) {$\frac16$};
	\node at (-4.2,-0.5) {$-\frac5{18}\,\hat\lambda$};
	\node at (7.5,-0.5) {$\frac14+\frac29\,\hat\lambda$};
		\node at (3.75,-0.5) {$\frac16+\frac{\hat\lambda}{18}$};
	\draw[color=red,line width=1pt,decorate,decoration={zigzag,amplitude=2,segment length=8,post length=0}] (-4,-0.001) -- (-1.5,0);
	\draw[color=red,line width=1pt,decorate,decoration={zigzag,amplitude=2,segment length=8,post length=0}] (5,0) -- (7,0);
		\draw[color=red,line width=1pt,decorate,decoration={zigzag,amplitude=2,segment length=8,post length=0}] (2.8,0) -- (3.3,0);
	\draw[color=black,line width=1pt,<-] (-5.5,0.01) -- (-4,0.01);
	\draw[color=black,line width=1pt] (-1.5,0.01) -- (2.8,0.01);
		\draw[color=black,line width=1pt] (3.3,0.01) -- (5.,0.01);
	\draw[color=black,line width=1pt,->] (7,0.01) -- (8.8,0.01);
	\draw[color=black,line width=1pt] (-1.5,-0.15) -- (-1.5,0.15);
	\draw[color=black,line width=1pt] (-4,-0.15) -- (-4,0.15);
	\draw[color=black,line width=1pt] (5.,-0.15) -- (5.,0.15);
	\draw[color=black,line width=1pt] (7,-0.15) -- (7,0.15);
	\draw[color=black,line width=1pt] (2.8,-0.15) -- (2.8,0.15);
	\draw[color=black,line width=1pt] (3.3,-0.15) -- (3.3,0.15);
	\end{tikzpicture} 
	\caption{The ANEC and unitarity bounds exclude certain ranges of $\xi$, in red zig-zag lines, below $\xi=0$, above $\xi=1/6$, and above $\xi=1/4$, when $d=4$.}
	\label{ranges xi 4d}
\end{figure}

\paragraph{ANEC bounds.} Combining the ANEC constraints \eqref{cfa} with the unitarity constraint \eqref{a}, at  \(\mathcal O (\lambda^1)\),  we obtain the following allowed regions for $\xi$ at fixed $\hat\lambda$:
 \begin{equation} 
	\begin{aligned}\label{table12}  
		\xi\leq -\frac{5}{18}\,\hat\lambda,\qquad\qquad 0<\,\,\xi\,\,\leq\frac16,\qquad\qquad \frac16+\frac{\hat\lambda}{18}\leq\xi\leq\frac14,\qquad\qquad\xi\geq\frac14+ \frac29 \,\hat \lambda,
	\end{aligned}
\end{equation} 
where it has to be understood that all the $\hat\lambda$-dependent bounds are only up to $\mathcal O(\hat\lambda^2)$.
The bounds \eqref{table12} prescribe forbidden gaps in the allowed range of $\xi$ below $\xi=0$,  above $\xi=\frac16$, and above $\xi=\frac14$, see figure \ref{ranges xi 4d}.  The ANEC bounds appear therefore stronger than   unitarity in sme regions of parameter space.  As opposed to the generic $d$ case, at $d=4$ three forbidden regions  arise instead of just two (see figures \ref{ranges xi}, \ref{ranges xi 2d}); the reason for this is the $\xi$-dependence of the unitarity constraints in this case.

We can constrain   the ANEC expectation value with the unitarity constraint at \(\mathcal {O} (\hat\lambda^2)\).  We get
\begin{equation} 
	\begin{aligned}\label{table2}  
			\xi \leq - \frac{5  }{ 18}\,\hat\lambda,\qquad \qquad 0 \leq \xi \leq \frac{1}{4},\qquad\qquad \xi \,\,\geq\,\, \frac 14 + \frac29 \,\hat \lambda,
	\end{aligned}
\end{equation}   
where again it has to be understood that all the $\hat\lambda$-dependent bounds are only up to $\mathcal O(\hat\lambda^2)$.
These bounds look quite different from	\eqref{table12} due to the qualitative difference between the first and second order unitarity constraint. 
Once again the ANEC appears to be stronger than unitarity.
 		
\section{Conclusions and Outlook}

In this paper we have derived constraints following from the ANEC in the case of the \(\lambda \phi^4\) theory at first order in the coupling. We have considered a state generated by the stress tensor and evaluated the correlators in perturbation theory. We find nontrivial bounds for the parameters of the theory. The ANEC and unitarity constraints \eqref{cfg-2}, \eqref{norm} (and the equivalent at $d=4$, \eqref{cfa}) originally depend on the dimensionless ratio between the energy of the state $q$ and a reference   (renormalization) scale $\mu$, which we then fix to the value \(\mu=q\). This would constrain the renormalized parameters at such energy scale, although the running in the case $d=4$ is not   captured by the perturbative order at which we work.
 We observe that the values \(\xi=0,\frac14\) play a prominent role;  the significance of this is  not yet clear.
 
 As a technical result, we compute  the 2-point function of the stress tensor to second oder in the coupling. This allows us to identify a candidate for the \(\mathrm c\)-anomaly coefficient under deformations, in a complementary approach to that of \cite{Hathrell:1981zb}. Our method allows us to directly evaluate this term and it is therefore easier to extend to higher orders.
 
Since we are using perturbation theory in the coupling \(\lambda\) only, this analysis is exact in \(\xi\). However, being this a perturbative analysis, the inclusion of higher order effects can dramatically affect the results, as we demonstrated with the analysis of the unitarity constraint at $d=4$. 
The technical derivation of the correlators is solid, but the bounds derived stand on a less firm ground.

An immediate extension of the work presented here is the inclusion in the ANEC correlator of the contributions of order \(\mathcal O(\lambda^2)\). Diagrammatically they are much more complicated than those considered here, thus they require a more sophisticated  technology in order to evaluate them in a form that is useful for our purposes.
The inclusion of higher orders would allow one to consider the effect of the running of the couplings \(\lambda\) and \(\xi\), thereby providing us with a working example in which to understand the implications of the ANEC for renormalization group flows.  This could eventually lead to insights on the \(\mathrm a\)-theorem, providing, for example, an interpolating function in terms of the 3-point function of the stress  tensor.

As a final note, we observe that \(\lambda \phi^4\), not having a UV limit, escapes the proof of \cite{Hartman:2016lgu}. It would be interesting to understand how such proof might be generalised.

\subsection*{Acknowledgements}
We thank   Petr Kravchuk, Olaf Lechtenfeld, Pierpaolo Mastrolia, and Christian Schubert   for discussions. 
We thank Chris Herzog, Sameer Murthy, and Stefan Theisen for discussions and comments on the first draft of this work, as well as  Hadi Godazgar, with whom our explorations on the ANEC were initiated. The work of T.B. has been supported by STFC grants ST/P000258/1 and ST/T000759/1.
LC thanks the hospitality of the Theoretical Physics group at  King's College London, where part of the work was done.
TB thanks the Albert Einstein Institut  for their hospitality while the work was being completed.

\appendix 

\section{Conventions and formulae}
\label{app:formulae}

We use the Lorentzian $(-{,}+{,}+{,}+)$ signature.  We define momentum measure as
\begin{equation}
\dbar^d p = \frac{d^d p}{  (2\pi)^d}.
\end{equation}
Our momentum-space correlators correspond to
\begin{equation}\label{yaa}
\begin{aligned}
\braket{T_{\mu\nu} (x)\,T_{\alpha \beta} (y)} 
& = \int\!  \dbar^d p_1 \,\dbar^d p_2 \ e^{i p_1 x + i p_2 y} 
\ (2\pi)^{d}\,
			\delta^{(d)}[p_1 + p_2] \ \braket{T_{\mu\nu} (p_1)\,T_{\alpha \beta} (p_2)} ,
\\
\braket{T_{\mu\nu} (x)\, T_{\rho\sigma}(z) \, T_{\alpha \beta} (y)} 
& = \int\!  \dbar^d p_1 \,\dbar^d p_2 \,\dbar^d p_3 \ e^{i p_1 x + i p_2 z + ip_3 y} \times
\\[-0.5em]
&\qquad\qquad\qquad
	{}	\times
	\ (2\pi)^{d}\,	\delta^{(d)}[p_1 + p_2+ p_3] \ \braket{T_{\mu\nu} (p_1)\, \,  T_{\rho\sigma}(p_2) \, T_{\alpha \beta} (p_3)} ,
\end{aligned}
\end{equation}
which
are defined for conserved momenta \(\sum p_i =0\).

Two-propagator loop integrals are given by
\begin{equation}\label{abj}
I^d_{mn}(p)
= \int \!\dbar^dq  \frac{1}{[q^2 ]^m[(q-p)^2]^n }
= \frac{(p^2)^{d/2-m-n}}{(4\pi)^{d/2}} \frac{\Gamma[m+n-\frac12d ]\, \Gamma[\frac 12 d -m ]\, \Gamma[\frac 12 d -n ]}{\Gamma[d-m-n] \, \Gamma[m]\, \Gamma[n]} .
\end{equation}
Standard formulae to relate tensor to scalar integrals can be found e.g.\ in \cite{Casarin:2021fgd}.

\section{Calculation of the eye diagram}
\label{app:eye}

In this appendix we explain the calculation of the eye-diagram (diagram $(2)f$ in figure \ref{fig:TTdiags}) contribution to the stress-tensor two-point function.
The contribution is given by the integral
\begin{gather}\label{eye}
\braket{T_{\mu\nu}(p)T_{\alpha\beta}(-p)}_{\mathrm E (2)f}
=
\int \dbar^d k \,\, \dbar^d w \,\,\dbar^d q\,\,\frac{V_{\mu\nu}(k,p)\,V_{\alpha\beta}(q,p) }{k^2\,q^2\,w^2\,(k+p)^2\,(q+p)^2\, (w-k+q)^{2} }\,,
\\
V_{\alpha\nu }  (k,p) =
k_\alpha k_\nu - k_{  (\alpha }  p_{\nu ) }  + \xi\, p_\alpha p_\nu 
- \frac12 \delta_{\alpha\nu} ( k^2  - k\cdot p+2 \xi \,p^2 ) \,.
\end{gather}
The inner loop can be computed using \eqref{abj} and we are left with
\begin{equation}\label{uaa}
\begin{gathered}
\braket{T_{\mu\nu}(p)T_{\alpha\beta}(-p)}_{\mathrm E (2)f}=\frac{ 1}{(4\pi)^{d/2}} \frac{\Gamma[2-\frac12d ] \Gamma[\frac 12 d -1] \Gamma[\frac 12 d -1 ]}{\Gamma[d-2] }    I_{\mu\nu\alpha\beta}(p) ,
\\
I_{\mu\nu\alpha\beta}(p) =
\int \dbar^d k \,\,\dbar^d q\,\,\frac{V_{\mu\nu}(k,p)\,V_{\alpha\beta}(q,p) }{k^2\,q^2\,(k+p)^2\,(q+p)^2 [(k-q)^{2}]^{2-d/2}}.
\end{gathered}
\end{equation}
The tensorial integral has the general structure
\begin{equation}\label{uab}
\begin{aligned} 
I_{\mu\nu\alpha\beta}=& A_1\,\delta_{\mu\nu} \,\delta_{\alpha\beta}
+A_2 \left( \delta_{\mu\alpha}\,\delta_{\nu\beta}+\delta_{\mu\beta} \,\delta_{\nu\alpha}\right)
+B_1\left(\delta_{\alpha\beta} \,  p_\mu p_\nu+\delta_{\mu\nu}  p_\alpha p_\beta \right)\\
&\quad {}
+B_2\left(\delta_{\nu\beta} p_\mu p_\alpha +\delta_{\mu\beta} p_\nu p_\alpha +\delta_{\mu\alpha} p_\nu p_\beta+\delta_{\nu\alpha} p_\mu p_\beta \right)
+C\,p_\mu p_\nu p_\alpha p_\beta
\end{aligned}
\end{equation}
with scalar coefficients.
By contracting with \(\delta_{\mu\nu} \,\delta_{\rho\sigma}\), \(p_\mu p_\nu p_\rho p_\sigma\), {\(\ldots\)} we obtain a system of equations for the coefficients. In doing so, in the integrand numerator one obtains powers of the loop momenta, powers of the external momenta, and mixed terms that can be rewritten in terms of the previous two by completing the square. In turn, the contractions reduce to iterated two-propagator integrals which can be computed exactly using \eqref{abj}, or to combinations of the scalar integral
\begin{equation}\label{uac}
I^d_{\Delta}(p) \equiv
\int \dbar^d k \,\,\dbar^d q\,\,\frac{1}{k^2\,q^2\,(k+p)^2\,(q+p)^2 [(k-q)^{2}]^{\Delta}}
\end{equation}
with \(\Delta = 2-d/2\), \(1-d/2\) or \( - d/2\). This  scalar integral has been   computed for generic \(\Delta\) in terms of  Gamma and hypergeometric functions \cite{Chetyrkin:1980pr,Kotikov:1995cw}, and is given by
\begin{equation}\label{uad}
\begin{aligned}
I^d_{\Delta}(p)
&=(p^2)^{3d/2-6}\frac{2}{(4\pi)^d}\,\Gamma\big[\frac{d}{2}-1\big]\,\Gamma\big[\frac{d}{2}-\Delta-1\big]\,\Gamma\big[3-d+\Delta\big]\,\times\\[-0.5em]
&\hspace{35mm}\times\Bigg[
\frac{2\,\Gamma[\frac{d}{2}-1]\,\, 
 _3 {F}_2\left(\left.
		      \begin{smallmatrix}
		      1\; , \; 2-\frac d2 +\Delta\; , \;d-2
		      \\
		      3-\frac d2 +\Delta\; , \; 1+\Delta
		      \end{smallmatrix}
		      \right\rvert 1 \right)
}{(d-2\Delta-4)\,\Gamma[1+\Delta]\,\Gamma [\frac 32d -\Delta-4]}-\frac{\pi\,\cot[\pi(d-\Delta)]}{\Gamma [d-2]}
\Bigg].
\end{aligned}
\end{equation}
We finally get
{	
\small
\begin{equation*} 
\begin{aligned} 
A_1&=\frac{  d^2-2d - 2   - 8 ( d-2) ( d+1) \xi + 
 16 ( d^2-1)  \xi^2  }{16   (d^2-1)} p^4 I_{ 2-\frac{d}{2}}^d(p)
  \\& \quad
 -
 \frac{ p^2  }{2 (d-2) (d^2-1) } I_{1-\frac{d}{2}} ^d(p)
  -
  \frac{  1 }{2 (d-2) (d^2-1) } I_{ -\frac{d}{2}}^d(p)
 \\& \quad
 -
\left(
36 d^5-294 d^4+817 d^3-830 d^2+101 d+174
-32 (d-2) (d+1) (2 d-3) (3 d^2-15  d+19) \xi 
\right) \times 
\\
& \qquad\qquad \times
\frac{  \Gamma [4-\frac{3
   d}{2}] \, \Gamma  [\frac{d}{2}-1]^2\, \Gamma [d-3]}{12 (4\pi)^d (d+1)\, \Gamma [2-\frac{d}{2} ] \, \Gamma [2 d-1]} \left(p^2\right)^{\frac{3 d}{2}-4} 
\\
A_2&=
\frac{p^4}{16 (d^2-1)}  I_{2-\frac d 2} ^d(p)+ \frac{  p^2 }{4(d-2)(d+1)} I_{1-\frac d 2}^d(p)  + \frac{ 1 }{4(d-2)(d+1)} I_{-d/2}^d(p)
\\
&\quad
   -\frac{\left(85 d^3-480 d^2+853 d-486\right)  
   \Gamma  [3-\frac{3 d}{2}]\,
    \Gamma [\frac{d}{2}-1] ^2 \,\Gamma [d-3]}{4 (4\pi)^d(d+1) \,\Gamma[1-\frac{d}{2} ] \, \Gamma [2 d-1]}\left(p^2\right)^{\frac{3 d}{2}-4} 
\end{aligned}
\end{equation*}}
{	
	\small
	\begin{equation*} 
	\begin{aligned} 
B_1&=
-\frac{ d^2-2d - 2   - 8 ( d-2) ( d+1) \xi + 
 16 ( d^2-1)  \xi^2   }{16(d^2-1)}p^2  I_{2-\frac  d 2}^d(p) 
   \\
   &\quad
      +\frac{1}{2 (d-2) \left(d^2-1\right)}I_{1-\frac  d 2}^d(p)
      -\frac{1 }{2 (d-2) \left(d^2-1\right)  p^2 }I_{-\frac  d 2}^d(p)
\\
&\quad-
     \left(15 d^5-128 d^4+372 d^3-407 d^2+90 d+60-4 (d-2) (d+1) (2 d-3) (9 d^2-49d+68) \xi \right)\times
   \\
   & \qquad\qquad \times\frac{ \Gamma [1-\frac{d}{2}]^3 \,\Gamma[\frac{d}{2}]^3 }{
         2^{3d-2} (4\pi)^{d-1}\Gamma [4-d]\,  \Gamma [2-\frac{d}{2}]^2 \,\Gamma[d-\frac{1}{2} ]\,
         \Gamma [\frac{d+3}{2}  ]\, \Gamma  [\frac{3 d}{2}-2] \, \big[ 2 \cos \frac{d \pi}{2}+\cos \frac{3 d \pi}{2}\big] }
         (p^2)^{\frac{3 d}{2}-5} 
         \\
B_2&=
 -    \frac{p^2 }{16(d^2-1)}I^d_{2-\frac{d}{2}} -\frac{1}{4(d+1)( d-2)}I_{1-\frac{d}{2}}^d(p)+\frac{1}{4(d-2)
      (d+1) p^2} I _{-\frac{d}{2}}^d(p)
   \\
   &\quad
-
\frac{  (d^4+80 d^3-475 d^2+858 d-492)  \, \Gamma [1-\frac{3 d}{2}] \, \Gamma [1-d]\,  \Gamma [d] \, \Gamma [\frac{3 d}{2}]\, \Gamma[1-\frac d2]\, \Gamma[\frac d2] }
   { 2^{3+3d}(4\pi )^{d-1}\Gamma [4-d] \,  \Gamma [2-\frac{d}{2}]^2 \,   \Gamma    [d-\frac{1}{2}] \,\Gamma [\frac{d+3}{2}]\, \Gamma  [\frac{3 d}{2}-2]}
   \left(p^2\right)^{\frac{3 d}{2}-5}
   \\
C&=
\frac{  d(d - 2)  - 8 ( d-2) ( d+1) \xi + 
 16 ( d^2-1)  \xi^2 }{16  (d^2-1 )  } I_ {2-\frac{d}{2}} ^d(p)
+ 
\frac{1 }{2  (d^2-1 ) p^2}I_{1-\frac{d}{2}}^d(p)
-
\frac{1}{2  (d^2-1 ) p^4}I_{-\frac{d}{2}}^d(p)
\\
&
\quad 
+  \left(21 d^4-152 d^3+477 d^2-774 d+480-16 (d-3) (d+1) (2 d-3) (3 d-10)\xi\right) \times
  \\
  &\qquad\qquad
  \times \frac{ \Gamma[1-\frac{3 d}{2}] \Gamma [1-d]\, \Gamma [1-\frac{d}{2}] \,\Gamma [\frac{d}{2}] \, \Gamma [d] \,\Gamma [\frac{3 d}{2}]}
     {3\,2^{ 3 d+1}  (4\pi)^{d-1} \Gamma [4-d] \, \Gamma[2-\frac{d}{2}] ^2\, \Gamma [d-\frac{1}{2}] \, \Gamma  [\frac{d+3}{2}]  \, \Gamma  [\frac{3 d}{2}-3]}\left(p^2\right)^{\frac{3 d}{2}-6} 
\end{aligned}
\end{equation*}}
To compute the norm \eqref{zba} we only need the coefficients $A_1$ and $A_2$, since we multiply the 2-point function by the transverse polarization tensors, but we give the full result for completeness.

\section{Full Euclidean stress-tensor 2-point function to order \( \mathcal O (\lambda^2)\)}
\label{app:full2pt}
In this appendix we write the contributions of all the diagrams up to 2-loop (figure \ref{fig:TTdiags}),  of the 2-point function, with the complete momentum dependence:
{ 
\begin{equation*}
\begin{aligned}
              \braket{T_{\mu\nu}  T_{\alpha\beta} }_{\mathrm{\E}(0)} & =  
           \frac{\Gamma[2-d/2] \, \Gamma[d/2-1]^2}{8(4\pi)^{d/2} (d^2-1) \,\Gamma[d-2]}\frac{1}{(p^2)^{2-d/2} }  
                       \times\Big[  (\delta_{\alpha\nu} \delta_{\beta\mu} + \delta_{\alpha\mu} \delta_{\beta \nu })  p^4
    	            \\
    	         & {}      \qquad
    	                 +\delta_{\alpha\beta} \delta_{\mu \nu}  p^4
    	                \left( d^2 (1-4 \xi )^2+d (8 \xi -2)-2 \left(8 \xi ^2-8 \xi +1\right)\right)
      	            \\
      	         & {}      \qquad
      	                 + p_{\alpha} p_{\beta} p_{\mu} p_{\nu} \left(d^2 (1-4 \xi )^2+d (8 \xi -2)-16 (\xi -1) \xi\right)
      	            \\
      	         & {}      \qquad
      	                 - p^2  \left( \delta_{\alpha \mu} p_{\beta} p_{\nu} + \delta_{\beta\nu} p_{\alpha}p_{\nu} +\delta_{\alpha\nu}p_{\beta}p_{\mu} +\delta_{\beta\mu}p_{\alpha} p_{\nu} \right)
      	            \\
      	         & {}      \qquad
      	         -p^2 ( \delta_{\mu \nu  } p_{\alpha} p_{\beta} + \delta_{\alpha\beta} p_{\mu}p_{\nu} )
      	         \left(d^2 (1-4 \xi )^2+d (8 \xi -2)-2 \left(8 \xi ^2-8 \xi +1\right)\right) 
      	         \Big]\,,\\
    \braket{T_{\mu\nu} T_{\alpha\beta} }_{\mathrm{\E}(1)}  
        & = 
      -      \left(\xi - \xi_d \right)^2 
       \frac{\Gamma[2-d/2]^2 \, \Gamma[d/2-1]^4}{(4\pi)^d\, \Gamma[d-2]^2}  
                 \frac{	 \left(p^2 \delta_{\alpha\beta} - p_\alpha p_\beta\right) 
                                                    	  \left(p^2 \delta_{\mu\nu} - p_\mu p_\nu \right)  
                       }{(p^2)^{4-d}}    \,,
\end{aligned}
\end{equation*} }

{ 
	\begin{equation*}
	\begin{aligned}
   \braket{T_{\mu\nu} T_{\alpha\beta} }_{\mathrm{\E}(2)a}  
       & =  
       -\frac{ \Gamma [ 7-\frac{3  {d}}{2} ] \, \Gamma  [\frac{ {d}}{2}-1 ]^4  }
           {9 ( 4 \pi)^{ 3  {d}/2}  ( {d}-4) (3  {d}-10) (3  {d}-8) \Gamma [2 d-4]
           }
           \frac{1}{ (p^2)^{4- 3d/2 }   } \delta_{\mu\nu} \delta_{\alpha\beta}  \,,
\\
    \braket{T_{\mu\nu} T_{\alpha\beta} }_{\mathrm{\E} (2)b}  
        & =  \frac{ \Gamma [4-\frac{3 d}{2}] \Gamma [\frac{d}{2}-1]^4  \left[(2d-5) (8\xi-1) p^2
                              \delta_{\alpha\beta}+p_\alpha p_\beta(d (3-16 \xi)+40\xi-8)\right]}
           {(4\pi)^{3 d/2} 
           24 (d-3) \Gamma (2 d-4)
                                (p^2)^{5-\frac{3  }{2}d}}
            \delta_{\mu\nu},
\\
    \braket{T_{\mu\nu} T_{\alpha\beta} }_{\mathrm{\E}(2)c}  
        & =    \braket{T_{\alpha\beta} T_{\mu\nu} (p \to -p) }_{\mathrm{\E} (2)b}  ,
\\
    \braket{T_{\mu\nu} T_{\alpha\beta} }_{\mathrm{\E}(2)d}  
        &=  \frac{(d-2) \Gamma  [7-\frac{3 d}{2}]  \Gamma[\frac{d}{2}-1]^4  }{({4\pi})^{3 d/2}\,9 (d-4)^2 (d-3) (3 d-10) (3 d-8) \Gamma [ 2 d-2]}
      \frac{1}{(p^2 )^{6-\frac{3 d}{2} }} \times
      \\&\hspace{-20mm} 
      \times 
         \Big[
       p^2\left( p_{\beta } p_{\mu }  \delta _{\alpha  \nu }  +
                 p_{\alpha } p_{\mu }  \delta _{\beta  \nu } +p_{\alpha } p_{\nu }  \delta _{\beta  \mu }+ p_{\beta } p_{\nu }  \delta _{\alpha  \mu } \right)  (d^2-6 d+12) 
            - \left(\delta _{\alpha  \nu } \delta _{\beta  \mu } +\delta _{\alpha  \mu } \delta _{\beta  \nu }\right) p^4 d
       \\ &\hspace{-15mm} 
 -   \delta _{\alpha  \beta } \delta _{\mu  \nu } p^4  \,	64 (d-3) (2 d-5) (2 d-3) 
 \left(\xi ^2-\frac{  (d-2)  }{2(2d-5)}  \xi +\frac{4 d^3-22 d^2+33 d-12}{ 64 (d-3) (2 d-5) (2 d-3) } \right) 
\\&\hspace{-15mm} 
-p_{\alpha } p_{\beta } p_{\mu } p_{\nu }
64 (d-3) (2 d-5) (2 d-3) \left( \xi ^2-\frac{ (d-2)   (3 d-10) }{4 (d-3) (2 d-5) } \xi -\frac{d ( 3 d-10) (  3 d-8)}{64 (d-3) (2 d-5) (2 d-3)} \right) 
  \\&\hspace{-15mm} 
   + p^2\left( p_{\alpha } p_{\beta }  \delta _{\mu  \nu }+p_{\mu } p_{\nu }  \delta _{\alpha  \beta }\right)
    64 (d-3) (2 d-5) (2 d-3)
    \times
    \\ &\times \left( \xi ^2-\frac{(d-2)(5 d-16)}{8 (d-3) (2 d-5) }  \xi +\frac{3 (d-2) (d^2-4  d+2)}{32 (d-3) (2 d-5) (2 d-3)} \right) 
         \Big],
\\[1em]
    \braket{T_{\mu\nu} T_{\alpha\beta} }_{\mathrm{\E}(2)e}  
& =  
- \left(\xi -\xi _d\right)^2 \frac{2^{5-3 d} \Gamma  [1-\frac{d}{2}] ^3 \Gamma [\frac{d}{2} ]^3
    }{
     (4\pi)^{\frac{3  }{2}d-\frac{3}{2}}  \Gamma [\frac{d-1}{2}]^3}
         \frac{	 \left(p^2 \delta_{\alpha\beta} - p_\alpha p_\beta\right) 
                                                         	  \left(p^2 \delta_{\mu\nu} - p_\mu p_\nu \right)  
                            }{(p^2)^{6-\frac{3  }{2}d}} ,
\end{aligned}
\end{equation*} }
with 
\(    \braket{T_{\mu\nu} T_{\alpha\beta} }_{\mathrm{\E}(2)f}  \)
  given in the previous appendix.

\section{Wightman functions in momentum space}\label{app:WR}

In this appendix we give some more detail on how we obtain the Wightman 2- and 3-point functions in momentum space from the corresponding Euclidean expressions, by means of a Wick rotation inside the Fourier transform. A complete discussion can be found in \cite{Bautista:2019qxj,Casarin:2021fgd}. 

The starting point is the position space expression of a Wightman function, which follows from its Euclidean expression by Wick rotating the time coordinate and using the $i\epsilon$ prescription: Lorentzian time of operators to the left in the correlator get a more negative imaginary part. 
The prescriptions for the 2- and 3-point functions are given in \eqref{acd} and \eqref{aia}.
The idea is then to implement this prescription in the  Euclidean Fourier representation \eqref{yaa} of the correlator, and further complexify the zeroth component of the momenta \(p_\mathrm {\E}^{\0} = i p^{\0} \) to obtain a Lorentzian Fourier kernel. In this procedure one has to deform the contour of integration in a way compatible with the limit \(\epsilon\to0\), which is thereby made manifest. The redefinition of the contour of integration depends on  the analytic properties of the integrand, i.e.\ the Euclidean expression of the momentum-space correlator. It is in them that the information on causality is encoded and becomes correspondigly reflected in the  resulting Lorentzian expression.   

Next we detail how we used this method to obtain the stress-tensor 2- and 3-point functions used in the paper.

\subsection*{2-point function}
The Euclidean correlator, given in appendix~\ref{app:full2pt},  has a simple power-law dependence on the external momentum, \((p^2_\mathrm E)^{-\alpha}\) with \(\alpha<1\). In this case the integrand presents a branch cut from \( p^{\0}_\mathrm E = +i |\vec p|\) to \(+i \infty \) and a symmetric one from \(p^{\0}_\mathrm E =- i  |\vec p|\) to \(-i \infty \), all branch points being simple  and the cuts going along the imaginary axis (additional positive integer powers of \(p^2\) do not change this fact).
The exponential in the \eqref{yaa} contains \(ip^{\0}_\mathrm E x^{\0}_\mathrm E = p^{\0}_\mathrm E (x^{\0} - i  \epsilon)\), \(\epsilon>0\), therefore in the complex \(p^{\0}_\mathrm E\) plane we close  the contour of integration in the upper half. The integral over the real \(p^{\0}_\mathrm E\) line can be therefore expressed in terms of an integral along the two sides of the upper branch cut multiplied by a factor \(\sim \sin \alpha\).  In terms of the Lorentian momentum \(p^{\0}=-ip^{\0}_\mathrm E \), this integral can be extended over the whole real domain introducing a step function. Overall, and reintroducing all factors, we obtain the formal substitution \((p^2_\mathrm E)^{-\alpha}  \to 2 \sin(\pi \alpha) \ \Theta[ p^{\0}-|\vec p\,|]\, |p^2|^{-\alpha} \) mentioned in the main text.\footnote{For  \(\alpha>1 \) and the branch points are not simple. Additional terms appear, see \cite{Casarin:2021fgd}.}

\subsection*{3-point function}
For the 3-point function the dependence on the external momenta is more complicated. The Euclidean expressions are given in \eqref{3ptE}.

We start with the free contribution, labelled as \((0)\).
We first consider \(p_{3\mathrm E}\), which gives simple poles in the complex \(p^{\0}_{3\mathrm E}\) plane for \(p^{\0}_{3\mathrm E} = - k^{\0}_\mathrm E \pm i |\vec p_3 + \vec k|\). The exponential contains \(- ip_{3\mathrm E}^{\0}z^{\0}_\mathrm E  =  p^{\0}_{3\mathrm E} (-z^{\0}+i\zeta)\), \(\zeta>0\), which forces   to close the contour of integration of \(p^{\0}_{3\mathrm E}\) on the lower half plane. The real \(p^{\0}_{3E}\) integral reduces to the residue of the pole with negative imaginary part.
We now turn to \(p_{1\mathrm E}\), which gives simple poles for \(p^{\0}_{1\mathrm E} =  k^{\0}_\mathrm E \pm i |\vec p_3 - \vec k|\). The exponential contains \( ip_{3\mathrm E}^{\0}( x^{\0}_\mathrm E-  z^{\0}_\mathrm E)  = p^{\0}_{3\mathrm E} (x^{\0}- z^{\0}- i( \epsilon-\zeta)) \), \(\epsilon > \zeta\), therefore we close the contour of \(p^{\0}_{1 \mathrm E} \) in the upper half plane and the integral reduces to the residue of the pole with positive imaginary part.
Finally we consider   \(k_\mathrm E\), which gives the simple poles \(k^{\0}_\mathrm E = \pm i |\vec k|\). \(k^{\0}_\mathrm  E\) appears in the exponential through the residues of the external momenta as \( i k^{\0}_\mathrm E \epsilon\), and since \(\epsilon>0\) the contour is closed on the upper half plane. The integral then is given by the residue on the pole \(k^{\0}_\mathrm E = + i |\vec k|\). Introducing the Lorentzian components \(p^{\0} = -i p^{\0}_\mathrm E\) for all three momenta we get the expression in  \eqref{aif}.

We now turn to the first order term labelled as \((1)a\). The other one, \((1)b\), is analogous.
We start considering \(p_{3\mathrm E}\), which goes like in the free case. Then, we consider the loop momentum \(k_\mathrm E\). We have simple poles at the values \(k^{\0}_{\mathrm E}= \pm i |\vec k|\) and \(k^{\0}_\mathrm E = p^{\0}_{1 \mathrm E} \pm i |\vec k - \vec p_1|\). \(k^{\0}_\mathrm E\) appears in the exponential through the value of \(p^{\0}_{3\mathrm E}\)  as \(i k^{\0}_\mathrm E \zeta \), therefore we close the contour of integration in the upper half plane and the integral reduce to the sum of the two residues with positive imaginary part. Finally we turn to \(p_{1 \mathrm E}\). Due to the factor \((p^2_{1\mathrm E})^{d/2-2}\), the zeroth component has a branch cut from \(p^{\0}_{1\mathrm E} = +i |\vec p_1|\) to \(+ i \infty \) and a symmetric one from \(p^{\0}_{1\mathrm E} = -i |\vec p_1|\) to \(- i \infty \), both  along the imaginary axis, as well as some simple poles. In the first term in the sum of the \(k^{\0}_\mathrm E\) residues, we have poles for \(p^{\0}_{1\mathrm E} = i(|\vec k |\pm | \vec k + \vec p_1|) \); the second term has simple poles for \(p^{\0}_{1\mathrm E} = i(-| \vec k + \vec p_1| \pm  |\vec k|) \). 
\(p^{\0}_{1 \mathrm E}\)  appears in the exponent respectively as \(ip^{\0}_{1 \mathrm E}( \epsilon - \zeta)\) and \( i p^{\0}_{1 \mathrm E} \epsilon\), therefore we   close both of them in the upper half plane. 
The integral over real \(p^{\0}_{1\mathrm E}\) is therefore expressed in terms of several contributions: the integral along the branch cut (multiplied by \(\sim \sin \frac{\pi d}2 \)), the isolated poles (giving a residue when they lie on the upper half plane) and a pole lying on the branch cut (which contributes with a residue multiplied by \(\cos  \frac{\pi d}2  \)). The contributions of the isolated poles cancel with each other. Rewriting in term of Lorentzian components for all the  momenta and extending it to an integral over the whole real line we finally obtain result is as in  \eqref{aif}.

\section{Massive free scalar}\label{app-mass}
In this appendix we inspect the ANEC in the case of the massive free boson   with generic conformal coupling $\xi$, whose mass term brakes scale invariance already at the classical level. The calculation proceeds as in the massless case, except for the fact that the propagator now exhibits the mass. We  only give the final expressions for the energy flux correlator,
\begin{equation}\label{kam}
   \begin{aligned}
     \braket{\mathcal E}
     & = \frac{ { q}^{d+1}\  \Theta[ q - 2 m]}{2^{d+1} {{(4  \pi)} }^{ d-2 }}
     				  \left( 1+\sqrt{r}\right)^{{d}-3} 
     				  \left|4 {\xi} ({\varepsilon_{11}}+{\varepsilon_{aa}})
     				  +r {\varepsilon_{11}} 
     				  	-\sqrt{r}
     				     ({\varepsilon_{11}}+{\varepsilon_{aa}}) \right|^2
   \\
   & \quad + \frac{ { q}^{d+1}\  \Theta[ q - 2 m]  }{2^{d+1} {{(4  \pi)} }^{ d-2 }}
   		 \left( 1 - \sqrt{r}\right)^{{d}-3} 
   		 \left| 	4 {\xi} ({\varepsilon_{11}}+{\varepsilon_{aa}})
      						+r{\varepsilon_{11}} 
      			+ \sqrt{r}
      			      ({\varepsilon_{11}}+{\varepsilon_{aa}}) 
      						\right|^2,
   \end{aligned}
\end{equation}
where we have defined \(r:= 1-4m^2/q^2\).
When \(m=0\) (\(r=1\)), only the first term in \eqref{kam} survives, and thus reproduces the result of the massless case, given by \eqref{cff} and coefficients \eqref{cfc} with $\lambda=0$. 
The condition \( q > 2m\) stems from the fact that the state under consideration,  generated by the stress tensor and therefore quadratic in the field, is a combination of two-particle states, which has a continuous spectrum above the mass threshold. This condition ensures that \eqref{kam} is real,  that $(1-\sqrt r)$ is non-negative, and therefore that the ANEC is satisfied.

\bibliographystyle{utphys}
\bibliography{LambdaPhi4}
	
\end{document}